\documentclass[letter,twocolumn,showpacs,aps,prb,floatfix]{revtex4}
\usepackage{graphicx,subfigure,epsfig,verbatim,psfrag,amsmath,amssymb,color}

\newcommand{\blue}[1]{{\textcolor{blue}{#1}}}
\newcommand{\mean}[1]{\left<#1\right>}
\newcommand{\abs}[1]{\left|#1\right|}
\makeatletter
\input epsf

\def\be{\begin{equation}}
\def\ee{\end{equation}}
\def\ba{\begin{eqnarray}}
\def\ea{\end{eqnarray}}
\makeatother

\begin{document}

\title{Thermodynamics of Ising spins on the triangular kagome lattice: Exact analytical method and Monte Carlo simulations}  
\author{Y.~L.~Loh, D.~X.~Yao,  and E.~W.~Carlson}
\affiliation{Department of Physics, Purdue University, West Lafayette, IN 47907}
\date{November 22, 2007}

\begin{abstract}

A new class of two-dimensional magnetic materials 
$\mbox{Cu}_{9}\mbox{X}_2(\mbox{cpa})_{6}\cdot x\mbox{H}_2\mbox{O}$
(cpa=2-carboxypentonic acid; X=F,Cl,Br)
was recently fabricated in which Cu sites form a Triangular Kagome Lattice (TKL).
As the simplest model of geometric frustration in such a system,
we study the thermodynamics of Ising spins on the TKL 
using exact analytic methods as well as Monte Carlo simulations.
We present the free energy, internal
energy, specific heat, entropy, sublattice magnetizations, and susceptibility.  We describe the rich  phase diagram of the model as a function of coupling constants, temperature, and applied magnetic field.
For frustrated interactions in the absence of applied field,
the ground state is a spin liquid phase with residual entropy per spin $s_0/k_B=\frac{1}{9} \ln 72\approx 0.4752\dots$.  
In 
weak 
applied field, the system maps to the dimer model on a honeycomb lattice, with residual entropy $0.0359$ per spin and quasi-long-range order with power-law spin-spin correlations that should be detectable by neutron scattering.
The power-law correlations become exponential at finite temperatures, but the correlation length may still be long.
\end{abstract}
\maketitle

\section{Introduction}
Geometrically frustrated spin systems 
give rise to many novel classical and quantum spin liquid phases.
They may have technological applications as refrigerants, via
adiabatic demagnetization,\cite{zhitomirsky-2003} in which reducing the applied magnetic field
results in a cooling effect as the spins absorb entropy from other degrees of freedom.
Unlike paramagnetic salts which are limited by ordering or spin-glass transitions due to residual interactions between the spins, geometrically frustrated systems  can remain in a disordered, cooperative paramagnetic state
down to the lowest temperatures.  
Furthermore, they may exhibit an enhanced magnetocaloric effect in the vicinity of phase transitions at finite 
applied fields.\cite{zhitomirsky-2004,derzhko-2004,isakov-2004,aoki-2004}

\begin{figure}
\resizebox*{0.65\columnwidth}{!}{\includegraphics{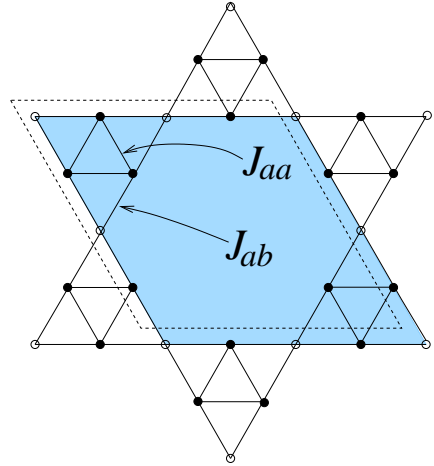}}
\caption{(Color online) The triangular kagome lattice (TKL).  Solid (open)
  circles represent ``a'' (``b'') sublattices.  Thick and thin lines
  represent interactions $J_{aa}$ and $J_{ab}$ respectively.  The shaded
  region represents a unit cell.  By shifting the cell slightly (as indicated by the dashed parallelogram) it can be seen that there are nine spins per unit cell.}
\label{lattice}
\end{figure}

A new class of two-dimensional magnetic materials
$\mbox{Cu}_{9}\mbox{X}_2(\mbox{cpa})_{6}\cdot x\mbox{H}_2\mbox{O}$
(cpa=2-carboxypentonic acid, a derivative of ascorbic acid; X=F,Cl,Br)
~\cite{gonzalez93,maruti94, mekata98} was recently fabricated, which is an
experimental realization of a new type of geometrically frustrated lattice.
The Cu spins in these materials are interconnected in a
``triangles-in-triangles'' kagome pattern, which we refer to as a triangular
kagome lattice (TKL, see Fig.~\ref{lattice}).

Experiments on the $\mbox{Cu}_{9}\mbox{X}_2(\mbox{cpa})_{6}\cdot x\mbox{H}_2\mbox{O}$
compounds show no spontaneous magnetization down to
at least $T=1.7$K,\cite{maruti94} consistent with a spin liquid ground state,
indicating that $J_{aa}$ is antiferromagnetic.
However, whether $J_{ab}$ is ferromagnetic or antiferromagnetic is still
an open question.   Based on the observation of a partial rather than saturated
magnetization,\cite{maruti94,mekata98}
Maruti {\em et al.} concluded that 
the intertrimer coupling is antiferromagnetic,  $J_{ab}<0$.\cite{maruti94}
In a later theoretical study using variational mean field theory
on the quantum Heisenberg model on the TKL,
Stre\v{c}ka\cite{strecka07} concluded that the intertrimer coupling is ferromagnetic, $J_{ab}>0$. 
Regardless, the lack of observed hysteresis despite the observation of 
a magnetization plateau in finite field\cite{maruti94}
is consistent with a multitude of ground states which 
can be connected by a series of local spin flips.   

In this paper we study the
classical TKL Ising model 
using exact analytic methods and Monte Carlo simulations.
The purpose of this study is to provide explicit predictions,
in order to determine to what 
extent the experiments can be explained in terms of a classical Ising model.
In particular, we find significant difference between the behavior of the
susceptibility as a function of temperature for $J_{ab}>0$ and $J_{ab}<0$,
and this may be used as an experimental means of distinguishing the two
cases.  Discrepancies between experiments and our theoretical predictions
will indicate the vector nature of the actual spins (XY or Heisenberg), the
effects of quantum fluctuations, or possibly higher order interactions.  This
model was previously studied by one of us \cite{daoxinMSthesis}.  In that
work, Monte Carlo simulations were used to study the phase
transitions and basic thermodynamics. 
Zheng and Sun\cite{zheng05} 
mapped the partition function to that of a kagome lattice and 
found an analytic expression for the phase boundary in zero field.

In this paper we present exact results in zero field at finite temperature,
and also in an applied field at zero temperature.
We report the full phase diagram as a 
function of coupling constants, temperature, and applied magnetic field.
We find several field-induced transitions.  
In particular, for frustrated interactions in applied field, we find a quasi-long-range ordered phase 
which maps to hard-core dimers on the honeycomb lattice. 
We complement these exact analytic results with Monte Carlo simulations
on the magnetization and susceptibility.  
We report the temperature-dependence of the magnetic susceptibility,
and show how it can be used to deduce the sign of the coupling constants.

The paper is organized as follows.  The TKL Ising model is described in
Section~\ref{model}.  In Section~\ref{zerofield}, we present exact results
for the TKL Ising model in zero field.  In Section~\ref{zerotemperature} we
present exact results at zero temperature.  In Section~\ref{finiteTfiniteh}
we present the phase diagram and describe the various phases.  In
Section~\ref{montecarlo} we present Monte Carlo results for the
susceptibility, spontaneous magnetization, and magnetization curves.  
In Section~\ref{discussion} we compare our model to
models of geometrically frustrated magnets on other lattices,
as well as to experiments on TKL systems.
In Section~\ref{conclusions} we present our conclusions, and in
Appendix
we present two mean-field approximations
in order to illustrate their failure in the frustrated regime.  

\section{Model \label{model}}

The TKL (Fig.~\ref{lattice}) can be obtained by 
inserting triangles inside the triangles of the kagome lattice,
for which it is sometimes referred to in the literature as the ``triangles-in-triangles''
kagome lattice.   
Alternatively, it can be
derived from the triangular lattice by periodically deleting seven out of
every sixteen lattice sites.  This structure has two different spin
sublattices ``a'' and ``b'', which correspond to small trimers and large
trimers, respectively. We study Ising spins on the TKL with two kinds of
nearest-neighbor interactions, the ``intratrimer'' couplings $J_{aa}$ and the
``intertrimer'' couplings $J_{ab}$.  Each spin has four nearest neighbors.
The Hamiltonian is
\begin{equation}
          H= -J_{aa}\sum_{i,j \in a} \sigma_{i}\sigma_{j} -
          J_{ab}\sum_{i\in a, j\in b} \sigma_{i}\sigma_{j} - h \sum_{i}
          \sigma_{i}
\label{e:hamiltonian}
\end{equation}
where $\sigma_{i}=\pm 1$, summations run over the nearest spin pairs
and all spin sites, $h$ is an external magnetic field.  
The shaded region in Fig.~\ref{lattice} is one unit cell, which contains 6
$a$-spins, 3 $b$-spins, 6 $a$--$a$ bonds, and 12 $a$--$b$ bonds.  We
shall use $N_{a}$ and $N_{b}$ to denote the total numbers of spins on
the a and b sublattices, so that $N_{a}:N_{b} =2:1$.  
The space group of the TKL is the same as that of the hexagonal lattice,
$p6m$, in Hermann-Mauguin notation. 

There are four energy scales in the problem: $J_{aa}$, $J_{ab}$, $T$, and $h$.  
We have found it most convenient to take $|J_{ab}|$ as the unit of energy.
Thus, the model can be described by a three-dimensional phase diagram in the space of the three dimensionless parameters 
$J_{aa}/|J_{ab}|$, $h/|J_{ab}|$, and $T/|J_{ab}|$.
The phase diagram also depends on the sign of $J_{ab}$.




\section{Exact results in zero field \label{zerofield}}
\begin{figure}[t]
\psfrag{J1}{$J_{aa}$}
\psfrag{J2}{$J_{ab}$}
\psfrag{J3}{$J_3$}
\psfrag{J4}{$J_4$}
\psfrag{J5}{$J_5$}
\psfrag{J6}{$J_6$}
\psfrag{J7}{$J_7$}
\psfrag{J8}{$J_8$}
\psfrag{J9}{$J_{h}$}
\psfrag{Nabla-Y}{$\nabla$--$Y$}
\centering
{\resizebox*{\columnwidth}{!}{\includegraphics{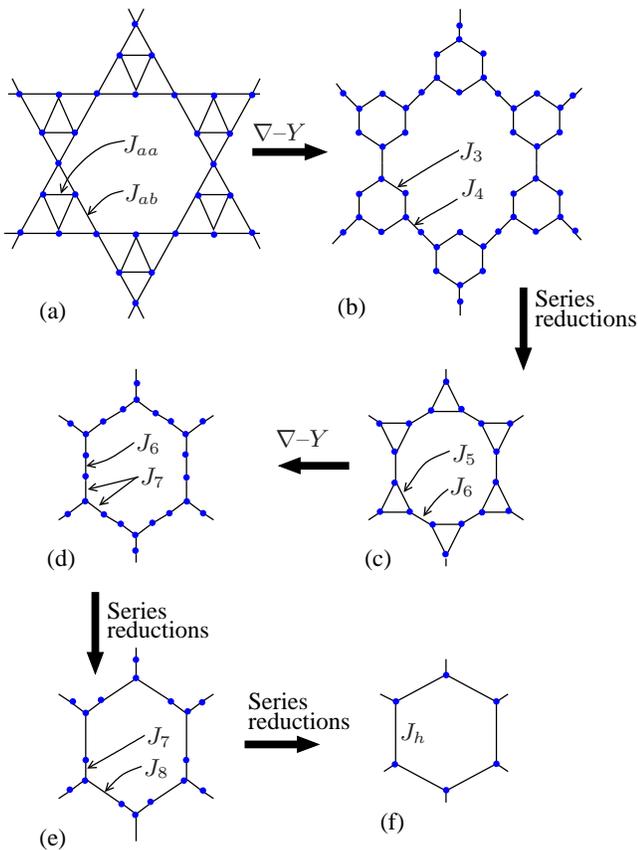}}}
\caption{(Color online). Transformation of a triangular kagome lattice (TKL) to a honeycomb lattice.  Figure (a) depicts a section of the TKL, with couplings $J_{aa}$ and $J_{ab}$.  The procedure begins by applying $\nabla$-$Y$ transformations to the six downward-pointing triangles in each unit cell.  This gives lattice (b).  Now take the two strong bonds ($J_3$) in series to give $J_5$, and the two weak bonds ($J_4$) in series to give $J_6$, to obtain a ``3--12 lattice'' (c).  Apply $\nabla$-$Y$ transformations to the triangles to obtain a decorated honeycomb lattice (d).  Finally, perform series reductions to obtain the honeycomb lattice (f) with a single coupling $J_h$.   
\label{f:tklreduction}}
\end{figure}

In this section we present exact analytic results for the TKL Ising model
in zero magnetic field ($h=0$), including the free energy,
internal energy, specific heat, and entropy.
We use a sequence of $\nabla$-$Y$
transformations and series reductions (shown in
Fig.~\ref{f:tklreduction}) to transform the Ising model on a TKL into
one on a honeycomb lattice, and then use the known solution for the
honeycomb lattice.  (See Ref.~\onlinecite{zheng05} for a similar
analysis that transforms the TKL to a kagome lattice keeping the
overall value of the partition function unchanged.)
We remark that frustrated Ising models with quenched bond disorder may
be studied by \emph{numerical} application of $\nabla$-$Y$ and
$Y$-$\nabla$ transformations\cite{loh2006,loh-jeremy2007} and/or Pfaffian
methods\cite{chineseremainder}.
Our results from this section are plotted in Figs.~\ref{f:fuctFM} and \ref{f:fuctAF}, using the analytic formulas below (as well as  series approximations for extreme values of $T$).   These thermodynamic quantities do not depend on the sign of $J_{ab}$.


\subsection{Effective coupling on the equivalent honeycomb lattice \label{sec:mapping}}

If an Ising model is
on a particular lattice
contains a spin $\sigma_0$ connected to only two other spins $\sigma_1$ and $\sigma_2$ via couplings $J_1$ and $J_2$, then we can `integrate out' $\sigma_0$ 
while preserving the value of the partition function, in order 
to obtain an effective coupling $J_{12}$. The transformation also produces a constant factor multiplying the partition function, $A$.  That is,
        \begin{align}
        \sum_{\sigma_0=-1}^{+1}
        e^{\beta (
                J_{1} \sigma_1 \sigma_0
        + J_{2} \sigma_2 \sigma_0
                )}
        &=      
        A e^{\beta 
                J_{12} \sigma_1 \sigma_2
                }
        \end{align}
for all combinations of values of $\sigma_1$ and $\sigma_2$.  
It is most convenient to write the effective coupling in terms of $t_{1},t_2,t_{12}$ where $t_{i}=\tanh \beta J_{i}$, and the partition function changes in terms of $x_{1},x_{2},x_{12}$ where $x_{i}=e^{-2\beta J_{i}}$:
        \begin{align}
        t_{12} &= t_1 t_2 \\
        A &= (1 + x_1 x_2) \sqrt{\frac{x_{12}}{x_1x_2}}
        \end{align}
The $x$'s and $t$'s are related by M\"obius duality transformations, $x_i=(1-t_i)/(1+t_i)$ and $t_i=(1-x_i)/(1+x_i)$.

Similarly, if a spin $\sigma_0$ is connected to only three other spins $\sigma_{1,2,3}$ via couplings $J_{1,2,3}$, we can `integrate out' $\sigma_0$ 
while preserving the partition function, in order
to obtain effective couplings $J_{23,31,12}$ together with a free energy shift.  This is known as a star-triangle or $Y$-$\nabla$ transformation.  The reverse transformation exists, and is known as a $\nabla$-$Y$ transformation: given a ``$\nabla$'' of couplings $J_{23,31,12}$, we can find an equivalent ``$Y$''.  
Again, it is convenient to use the variables $t_1=\tanh \beta J_1$
(and similarly for $t_2$, $t_3$) and $x_1=e^{-2 \beta J_1}$:
        \begin{equation}
        t_{1} = {\sqrt{a_1a_2a_3/a_0} \over a_1} \quad\text{(cycl.)} 
\end{equation}
where
        \begin{align}
        a_0 &= 1 + t_{23} t_{31} t_{12} \\
        a_1 &= t_{23} + t_{31} t_{12} \quad\text{(cycl.)} \\
        A &= 
                \frac{1}{1 + x_1x_2x_3} 
                        \sqrt{\frac{x_1x_2x_3}{x_{23}x_{31}x_{12}}}~,
        \end{align}
and ``cycl.'' means that $t_{2}$, $t_3$, $a_2$, and $a_3$ are found by cyclic permutation of the indices $1,2,3$.  In general the $a$'s may be negative or complex-valued, so 
that it is not correct to 
replace $\sqrt{a_1a_2a_3/a_0}/a_1$ by $\sqrt{a_2a_3/a_0a_1}$.

Using a sequence of $\nabla$-$Y$ transformations and series reductions, we transform the TKL Ising model (with couplings constants $J_{aa}$ and $J_{ab}$) into a honeycomb Ising model (with a single coupling constant $J_h$), as shown in Fig.~\ref{f:tklreduction}.  The transformation equations (in terms of the $t_i=\tanh \beta J_i$ variables) are:
        \begin{align}
        t_3     &= \sqrt{ (t_{aa} + {t_{ab}}^2) / (1 + t_{aa}{t_{ab}}^2)} \\
        t_4     &= (t_{ab}+t_{aa}t_{ab}) / \sqrt{ (t_{aa} + {t_{ab}}^2)(1 + t_{aa}{t_{ab}}^2) } \\
        t_5 &= {t_3}^2 \\
        t_6 &= {t_4}^2 \\
        t_7 &= 1 / \sqrt{ (t_5 + {t_5}^{-1} - 1)}  \\
        t_8 &= t_6 t_7 \\
        t_h &= t_8 t_7
        \end{align}
We can write $t_h$ directly in terms of $t_{aa}$ and $t_{ab}$:
        \begin{align}
        t_h &= 
        \frac{
                \left(1+{t_{aa}}\right)^2 {t_{ab}}^2
        }{
                \left(1-{t_{aa}}+{t_{aa}}^2\right)  \left(1+{t_{ab}}^4\right)
        -       \left(1-4{t_{aa}}+{t_{aa}}^2\right) {t_{ab}}^2  }
                                \label{th-taa-tab}
        \end{align}
It will be convenient to 
rewrite this in terms of $x_i = e^{-2 \beta J_{i}}$, as this is simpler:
\begin{align}
       x_h &=
\frac{2 \left(3 {x_{aa}}^2+1\right) {x_{ab}}^2}{{x_{ab}}^4+6 {x_{aa}}^2
   {x_{ab}}^2+1}
\label{xh-xaa-xab}
\end{align}

\begin{figure}
\psfrag{J1}{$J_1$}
\psfrag{J2}{$J_2$}
\psfrag{J3}{$J_3$}
\psfrag{J4}{$J_4$}
\psfrag{J5}{$J_5$}
\psfrag{J6}{$J_6$}
\psfrag{J7}{$J_7$}
\psfrag{J8}{$J_8$}
\psfrag{J9}{$J_h$}
\centering
{\resizebox*{0.47\textwidth}{!}{\includegraphics{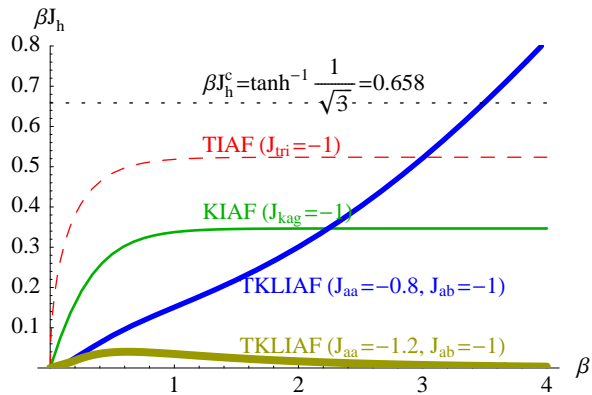}}}
\caption{(Color online). Effective dimensionless coupling $\beta J_h$ of the honeycomb lattice Ising model as a function of the original inverse temperature $\beta$, for the triangular Ising antiferromagnet (TIAF), kagome Ising antiferromagnet (KIAF), or TKL Ising antiferromagnet (TKLIAF).
In the case of the TIAF, $\beta J_h$ is imaginary, and $\text{Im~} \beta J_h$ is shown as a dashed curve; for the KIAF and TKLIAF, $\beta J_h$ is real.
For the TKLIAF in the unfrustrated regime, $\beta J_h$ crosses the dotted line (the critical coupling of the honeycomb Ising model), indicating a phase transition.  The other models do not have phase transitions.
\label{f:effhoneycoupling}}
\end{figure}

For Ising models, 
both the triangular lattice and the kagome lattice can be transformed into
the honeycomb lattice by a similar procedure.  Figure
\ref{f:effhoneycoupling} shows the effective dimensionless coupling of the
honeycomb lattice, $\beta J_h$, as a function of the inverse temperature for
antiferromagnetic Ising models
on different lattices for comparison.  Note that:  

\begin{itemize}
\item For the triangular Ising AF (TIAF), the effective honeycomb coupling is imaginary, and there is no long-range order.  
\item For the kagome Ising AF (KIAF), the effective honeycomb coupling is real.  As the kagome couplings are increased, the effective honeycomb coupling increases
until it saturates at large $\beta$.
However, it never grows beyond the critical coupling for the honeycomb model, $\beta J_h^c = \tanh^{-1} 1/\sqrt{3} \approx 0.658$.  Therefore the 
KIAF does not have a phase transition.  
\item For the triangular kagome lattice Ising antiferromagnet (TKLIAF) in the unfrustrated regime (e.g., $J_{aa}=-0.8, J_{ab}=-1$), the plot of $\beta J_h$ intersects the dotted line, indicating a phase transition at $\beta \approx 3.5$.
\item In contrast, for the TKLIAF in the frustrated regime ($J_{aa}=-0.8, J_{ab}=-1$), $\beta J_h$ grows to a maximum and then decays to zero, indicating the absence of a phase transition.  Paradoxically, 
and unlike the other lattices,
\emph{stronger} bare couplings lead to a \emph{weaker} effective coupling!  The fact that $\beta J_h\rightarrow 0$ as $\beta\rightarrow \infty$ is closely connected to the fact that the residual entropy of the TKL lattice has the simple value of $\ln 72$ per unit cell (as we will show), unlike the cases of the TIAF\cite{wannier50} and KIAF\cite{kano53} in which the residual entropies are non-trivial two-dimensional integrals.
\end{itemize}

\subsection{Phase boundary}


The phase boundary of the TKL Ising model in zero applied field was
calculated in Ref.~\onlinecite{zheng05}.  We show an alternative exact
derivation of the results here for pedagogical reasons.  Once we have used
the techniques of Sec.~\ref{sec:mapping} to map the TKL Ising model into the
Ising model on a honeycomb lattice, we can use known results for the
honeycomb lattice.  The critical temperature of the honeycomb Ising model is
given by $t^c_h = \tanh \beta J_h = 1/\sqrt{3}$, or, equivalently, $x^c_h =
\exp -2\beta J_h = 2 - \sqrt{3}$.  Substituting in the equivalent coupling of
the honeycomb lattice, Eq.~\ref{xh-xaa-xab}, leads to an implicit equation
for the critical temperature $1/\beta_c$ of the TKL Ising model:  
  \begin{align}
       e^{-4\beta_c J_{aa}} &= 
                                        (\sqrt{3}-1) \cosh 4\beta_cJ_{ab}
                                        -\left(\sqrt{3} + 1 \right)     .
        \label{e:phaseboundary}
   \end{align}
Eqn.~7 of Ref.~\onlinecite{zheng05} is equivalent to the simpler expression reported here. 
This critical curve is plotted in Fig.~\ref{zerofieldphasediagram}. 
For large ferromagnetic $J_{aa}$, the critical temperature saturates at a finite value,
$T_c/|J_{ab}| \approx
4/\ln\left( 2+\sqrt{3}+\sqrt{6+4\sqrt{3}} \right) \approx 2.00838$.
As $J_{aa}$ is reduced towards $-1$, the critical temperature falls to zero.  Near 
$J_{aa}=-|J_{ab}|$, 
the critical curve is approximately linear:

\begin{figure}[!t]
\centering
{\resizebox*{0.47\textwidth}{!}{\includegraphics{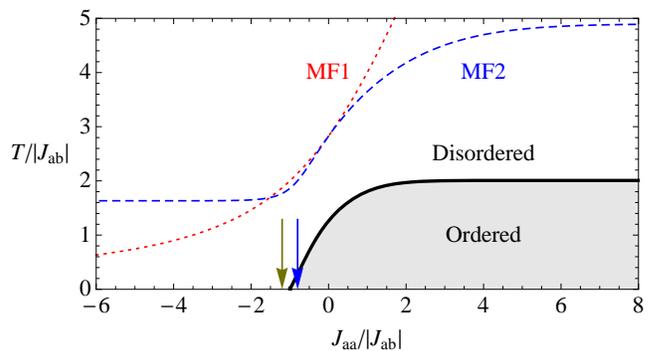}}}
\caption{(Color online) Phase diagram of the TKL Ising model in the $(J_{aa},T)$ plane, for $J_{ab}=\pm 1$ and $h=0$.  The thick curve is the exact solution (equivalent to that in Ref.~\onlinecite{zheng05}).  The dotted and dashed curves are mean-field approximations (see Appendix, Sec.~\ref{meanfield}).  
The ordered phase is ferromagnetic if $J_{ab}>0$ and ferrimagnetic if $J_{ab}<0$.   The disordered state is paramagnetic.  For $J_{aa} < -J_{ab}$, it persists down to $T=0$, where the entropy is $\ln 72$ per unit cell and the susceptibility obeys a Curie law.  The arrows indicate the two TKLIAF cases discussed in Fig.~\ref{f:effhoneycoupling}.   
\label{zerofieldphasediagram}}
\end{figure}

Since the mapping from the TKL to the honeycomb lattice preserves the nature of the singularity in the partition function, the phase transition is a continuous second-order transition in the 2D Ising universality class.

In zero field ($h=0$) the partition function $Z$ is invariant under a change of sign of $J_{ab}$, and the topology of the phase diagram is independent of this sign, although the identification of the phases is not.
First consider the model for the case $J_{aa}=0$, where the TKL reduces to a decorated kagome lattice. If $J_{ab}$ is ferromagnetic, the model develops {\em ferromagnetic} order below the Curie temperature.
If $J_{ab}$ is antiferromagnetic, the model develops {\em ferrimagnetic} order
below the ordering temperature.
Although the decorated kagome lattice is bipartite,
the numbers of spins on the $a$- and $b$-sublattices are not equal:  there are twice as many $a$-spins as $b$-spins.  Hence we have \emph{ferrimagnetic} order, with unequal numbers of up and down spins producing a net moment.  
However, 
the transition temperature, free ener:gy, internal energy, specific heat, and entropy
are independent of the sign of $J_{ab}$ in the absence of applied field.
Now introduce the coupling $J_{aa}$.  If this is ferromagnetic, it has very little effect, since the $a$-spins already have a tendency to align.  However, if $J_{aa}$ is antiferromagnetic, it fights against the ordering induced by $J_{ab}$.  If $J_{aa}$ is 
sufficiently antiferromagnetic, ($J_{aa}/|J_{ab}| < -1$), the system is in a frustrated regime with no order even at zero temperature.

\subsection{Partition function}

\newcommand{\shifted}{\text{shifted}} 

The partition function 
per unit cell, $z_{\rm TKL}$, 
of the TKL Ising model is equal to that of the equivalent honeycomb lattice 
$z_H$,
multiplied by the factors $z_1, \dotsc, z_6$ below
which are accumulated during the sequence of $\nabla$-$Y$ transformations and series reductions
necessary to effect the transformation.
These factors are
\begin{align}
z_1 &=
        \frac{ 1 }{ 1 + x_4 x_3{}^2}    
        \sqrt{\frac{ x_4 x_3{}^2 }{x_{aa} x_{ab}{}^2 }} 
\quad\text{(first $\nabla-Y$)}\\
z_2 &=
  (1 + x_3{}^2)
        \sqrt{\frac{x_5}{x_3{}^2}} 
\qquad\text{($J_3$ in series)} \\
z_3 &=
  (1 + x_4{}^2)
        \sqrt{\frac{x_6}{x_4{}^2}} 
\qquad\text{($J_4$ in series)} \\
z_4 &=
        \frac{ 1 }{ 1 + x_7{}^3}    
        \sqrt{\frac{ x_7{}^3 }{ x_5{}^3 }} 
\qquad\text{(second $\nabla-Y$)} \\
z_5 &=
        \frac{ 1 }{ 1 + x_6 x_7 }    
        \sqrt{\frac{ x_8 }{ x_6 x_7 }} 
\qquad\text{($J_6,J_7$ in series)} \\
z_6 &=
        \frac{ 1 }{ 1 + x_8 x_7 }    
        \sqrt{\frac{ x_9 }{ x_8 x_7 }} 
\qquad\text{($J_8,J_7$ in series)} 
\end{align}
The total accumulated partition function change is therefore
        \begin{align}
  z_{\rm TKL}
  &=z_1{}^6 z_2{}^6 z_3{}^3 z_4{}^2 z_5{}^3 z_6{}^3~z_H
        \nonumber\\
        &=
        \frac{
                (1+x_{ab}{}^2) ^2  
                (1 + 6x_{aa}{}^2x_{ab}{}^2 + x_{ab}{}^4) ^3 
        }{
                1 + 2(1 + 6x_{aa}{}^2)x_{ab}{}^2 + x_{ab}{}^4
        }~z_H~.
        \end{align}
The partition function per unit cell of the honeycomb lattice has been calculated in the literature by, {\em e.g.}, the Pfaffian method\cite{kasteleyn1963,fisher1966}.  It is
\begin{align}
  z_H(x_h)
  &=\frac{\sqrt{2} (1 - {x_h}^2)}{x_h} 
    \exp \left[ \tfrac{1}2 \Omega(w(x_h)) \right]
\end{align}
where
 \begin{align}
\Omega(w) 
 &=\int_0^{2\pi} \frac{dp}{2\pi}  \int_0^{2\pi} \frac{dq}{2\pi}
   \ln (w - \cos p - \cos q - \cos (p+q))
 \label{Omega-w}
 \end{align}
and
\begin{align}
    w(x_h)
  &= \frac{1 - 2{x_h} + 6{x_h}^2 - 2{x_h}^3 + {x_h}^4}{2x_h(1-x_h)^2}
.
  \label{w-xh}
\end{align}

For the purposes of numerical evaluation,
we rewrite the function $\Omega(w)$ in the following form
\begin{align}
\Omega(w) 
&=\frac{2}{\pi} \int_0^{\pi/2} dp
  \ln \left[\cos p + \text{arccosh}  \frac{w-\cos 2p}{2 \cos p}
  \right]~.
\end{align}
In order to get accurate numerical results one has to further split the range
of integration according to the singularities of the integrand.
   
Thus the partition function of the TKL Ising model (per unit cell) is
\begin{align}
z_{\rm TKL}(x_{aa},x_{ab})
&=\Psi(x_{aa},x_{ab}) 
        \exp \left[ \tfrac{1}2 \Omega (w (x_h (x_{aa}, x_{ab}))) \right]
\end{align}
where
        \begin{align}
        \Psi
        &=
        2 {x_{aa}}^{-3} {x_{ab}}^{-5}  
           \left(1 - {x_{ab}}^4\right)^2 
\times\nonumber\\&{}\quad
                \sqrt{ \left( 1 + 3 {x_{aa}}^2 \right) 
                                        \left( 1 + 6 {x_{aa}}^2 {x_{ab}}^2 +  {x_{ab}}^4 \right) }
\label{eqn:Psi}
        \end{align}
and $\Omega$, $w$, and $x_h$ are defined in Eqs.~(\eqref{Omega-w}), (\eqref{w-xh}), (\eqref{xh-xaa-xab}),
respectively.
The total partition function $Z_{\rm TKL}$ is related to the partition function per unit cell $z_{\rm TKL}$
by $Z_{\rm TKL} \equiv z_{\rm TKL}^N$, where N is the number of unit cells.
We show plots of 
$-{\rm ln}z_{\rm TKL}/(\beta |J_{aa}|)$
 in Figs.~\ref{f:fuctFM} and \ref{f:fuctAF} (red curves)
in the unfrustrated and frustrated regimes, respectively. 


\subsection{Energy}
The energy per unit cell of the TKL Ising model can be obtained by differentiation of the 
partition function:
        \begin{align}
        u
        &=-\frac{d\ln z}{d \beta}
        =
        -\frac{d x_{aa}}{d \beta} \frac{\partial \ln z}{\partial x_{aa}}
        -\frac{d x_{ab}}{d \beta} \frac{\partial \ln z}{\partial x_{ab}}
\\
        &=\sum_{i=aa,ab}
                J_i x_i \left[2\frac{\partial \ln\Psi}{\partial x_i}
                + \frac{\partial x_h}{\partial x_i} \frac{d w}{d x_h} \frac{d \Omega}{d w}
                \right]~,
        \end{align}
        where $\Psi$ is given in Eqn.~\ref{eqn:Psi}.  $\frac{d \Omega}{d w}$
        is the Green function of a particle on a triangular lattice and can
        be expressed in terms of the complete elliptic integral of the first
        kind, $K$:\cite{horiguchi1992}
        \begin{align}
        \frac{d \Omega}{d w}
        &=\int_0^{2\pi} \frac{dp}{2\pi}  \int_0^{2\pi} \frac{dq}{2\pi}
                \frac{1}{ w - \cos p - \cos q - \cos (p+q) }
\\      
        &=-\tfrac2{\pi
   (-w-1)^{3/4}(-w+3)^{1/4}} 
\times\\&{}\qquad\quad
                K
                \left(  \tfrac{1}2+
   \tfrac{w^2-3}{2(w+1)(-w-1)^{1/2}(-w+3)^{1/2}} \right).
        \end{align}
We show plots of $-u/|J_{aa}|$ in Figs.~\ref{f:fuctFM} and \ref{f:fuctAF} (green curves).

\subsection{Specific heat}
\begin{figure}[htb]
\centering
{\resizebox*{0.47\textwidth}{!}{\includegraphics{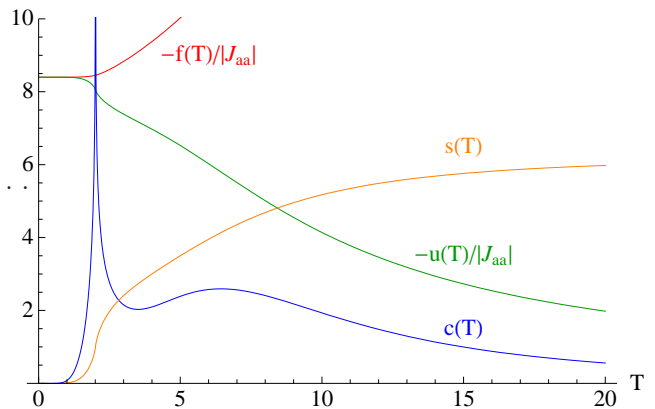}}}
\caption{(Color online). Thermodynamic functions vs temperature $T$ for unfrustrated (ferromagnetic) couplings $J_{aa}=5$ and $|J_{ab}|=1$.  Red: free energy $f(T)=-k_B T\ln Z$.  Green: energy $u(T)$.  Blue: heat capacity $c(T)$.  Yellow: entropy $s(T)$.  All values quoted per unit cell; each unit cell contains 9 sites.  
\label{f:fuctFM}}
\end{figure}

\begin{figure}[htb]
\centering
{\resizebox*{0.47\textwidth}{!}{\includegraphics{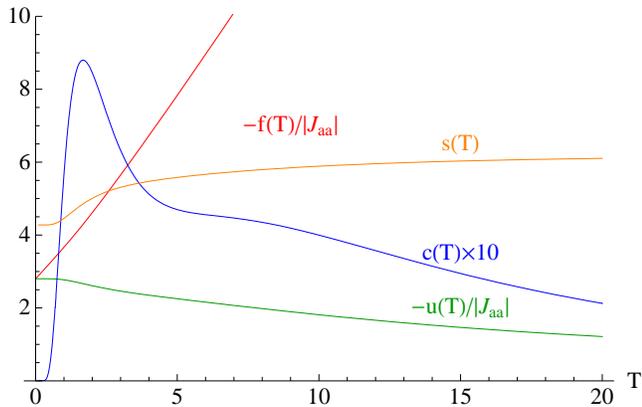}}}
\caption{(Color online). Thermodynamic functions vs temperature $T$ in frustrated regime, $J_{aa}=-5$ and $|J_{ab}|=1$, where the intratrimer coupling $J_{aa}$ is strong and antiferromagnetic. 
\label{f:fuctAF}}
\end{figure}

The heat capacity per unit cell, $c=\frac{du}{dT}$, can be obtained by further differentiation:
\newcommand{\dd}{\partial}
\begin{widetext}

\begin{align}
c &= 2\Bigg\{
(\beta J_{aa})^2 x_{aa} \left[  
        \frac{\dd^2\ln\Psi}{\dd x_{aa} \dd x_{aa}}
+ \left(  \frac{\dd x_h}{\dd x_{aa}}  
                + \frac{\dd^2 x_h}{\dd x_{aa}{}^2} x_{aa} 
        \right)
                w' \Omega'
+ \left( \frac{\dd x_h}{\dd x_{aa}} \right)^2 x_{aa}
                        (w'' \Omega' + (w') ^2 \Omega'')
\right]
\nonumber\\&{}~~~
+
\left[ \text{previous term with $aa$ replaced by $ab$} \right]
\nonumber\\&{}~~~
+
2\beta^2 J_{aa} J_{ab} x_{aa} x_{ab} \left[
                \frac{\dd^2\ln\Psi}{\dd x_{aa} \dd x_{ab}}
+ \frac{\dd^2 x_h}{\dd x_{aa} \dd x_{ab}} w' \Omega'
+ \frac{\dd x_h}{\dd x_{aa}} \frac{\dd x_h}{\dd x_{ab}} 
                        (w'' \Omega' + (w')^2 \Omega'')
\right]
\Bigg\}
\end{align}
\end{widetext}
where $w'$, $\Omega'$, etc., represent derivatives of the functions $w(x_h)$ and $\Omega(w)$ with respect to their arguments.
We show plots of $c$ in Figs.~\ref{f:fuctFM} and \ref{f:fuctAF} (blue curves).

In the unfrustrated case ($J_{aa}>0$, Fig.~\ref{f:fuctFM}), 
the specific heat has a broad hump just above $T = J_{aa}$,
and a sharp peak 
near $T = 2 |J_{ab}|$.
The broad hump is due to ferromagnetic alignment within each $a$-plaquette.
The sharp peak corresponds to the phase transition to a ferromagnetic (for $J_{ab}>0$) or ferrimagnetic (for $J_{ab}<0$) state, governed by the weakest links, $J_{ab}$.  
The position of the sharp peak is consistent with the 
solution of Eqn.~\ref{e:phaseboundary}.  (See also Fig.~\ref{zerofieldphasediagram}.)
In the frustrated case, 
broadened features remain at both of these energy scales,
as shown in Fig.~\ref{f:fuctAF}.

\subsection{Zero-temperature limit: residual entropy \label{s:entropy}}

Far from the critical curve, results for two limits can be obtained, corresponding to the
ordered phase and to the disordered phase (which persists even at zero temperature).

In the first case, $J_{aa}/|J_{ab}| > -1$, the system 
orders at low temperatures, going into either the ferromagnetic state (for $J_{ab}>0$) or ferrimagnetic state
(for $J_{ab}<0$).
In the low temperature limit, the partition function and internal energy may be expanded as 
        \begin{align}
        \ln Z(\beta) 
        &= 6 \ln x_{ab}  - 3 \ln x_{aa} + \frac{6 {x_{aa}}^2}{{x_{ab}}^2} + \dotso,
\\
        u(\beta)        
        &= -12|J_{ab}| + 6|J_{aa}|
        + 24 e^{-4\beta |J_{ab}-J_{aa}|} |J_{ab}-J_{aa}|
        + \dotso .
        \end{align}
As $T \rightarrow 0$, the residual entropy 
is zero, whether in the ferromagnetic phase or the ferrimagnetic phase. 

Suppose $J_{aa}/|J_{ab}| < -1$.  In this case, the model becomes frustrated
when $T\rightarrow 0$, $\beta \rightarrow \infty$.  We have $x_{aa}, x_{ab}
\rightarrow \infty$, $x_h \rightarrow 1^-$, $w \rightarrow \infty$.
Expanding $\Omega(w) \approx \ln w$ in a series in $w$, and then expanding
$\ln Z$ as a series in $x_{aa}$ and $x_{ab}$, we find:
        \begin{align}
        \ln Z(\beta) 
        &= \ln 72 + \ln x_{aa} + 2 \ln x_{ab}  
\nonumber\\&{}
        + \frac{2}{{x_{ab}}^2} + \frac{1}{3{x_{aa}}^2} + \frac{{x_{ab}}^2}{6{x_{aa}}^2} + \dotso .
        \end{align}
We can thus obtain the following low-temperature approximation for the energy per unit cell:
        \begin{align}           
        u(\beta)        &= -2|J_{aa}| - 4|J_{ab}|
        + \tfrac2{3} e^{-4\beta |J_{aa}-J_{ab}|} |J_{aa}-J_{ab}|
\nonumber\\&{}
        + \tfrac2{3} e^{-4\beta |J_{aa}+J_{ab}|} |J_{aa}+J_{ab}|
        + \tfrac{4}{3} e^{-4\beta |J_{aa}|} |J_{aa}|
\nonumber\\&{}
        + 8 e^{-4\beta |J_{ab}|} |J_{ab}|
        + \dotso .
        \end{align}
The first two terms are the ground state energy.
The other terms represent different types of excitatiotns about the ground
state(s).  The coefficients represent relative degeneracies of excited states and ground states, and the exponents represent excitation energies.

The first term in the series expansion of $\ln Z$ gives the residual entropy per unit cell:
        \begin{align}           
          s_0 &= \lim_{\beta J_{ab}\rightarrow -\infty} \lim_{\beta
            J_{aa}\rightarrow -\infty}
          \left( \ln Z + \beta u\right)
          = \ln 72.
        \end{align}
Thus the residual entropy is exactly $\ln 72 \approx 4.2767\dots$ per unit cell, or 
$\frac{1}{9} \ln 72  \approx 0.4752\dots$ per site.  This number will be discussed in more detail in Sec.~\ref{zerotemperature}.



\section{Exact results at zero temperature \label{zerotemperature}}

In this section we show how the zero-temperature phase diagram (along with the thermodynamic properties and correlations of the various phases) can be systematically deduced 
with and without applied field
by considering ground states of large triangular plaquettes.
By explicitly comparing the ground state energies, we
derive the phase diagram.  
The results of this section are summarized in 
Figs.~\ref{phase_diag_Jab-1_T0}  and \ref{phase_diag_Jab1_T0}.

\subsection{Zero Field (phases V and VI)\label{sec:zerofield}}

The phase diagram for zero applied field is shown in 
Fig.~\ref{zerofieldphasediagram}.
For $J_{aa}>-|J_{ab}|$ and low temperature, the system is in an ordered phase 
which is ferromagnetic for $J_{ab}>0$, and ferrimagnetic if $J_{ab}<0$.  
For $J_{aa}<-|J_{ab}|$, the system remains disordered even at zero temperature,
with a residual entropy of $s_0={\rm ln 72}$.
The zero-field disordered phase is labeled phase V in 
Figs.~\ref{phase_diag_Jab-1_T0}  and \ref{phase_diag_Jab1_T0},
and the zero-field ordered phase is labeled phase VI.

\begin{figure}[htb]
{\centering
\subfigure[All three $b$-spins pointing up]
{\resizebox*{0.9\columnwidth}{!}{\includegraphics{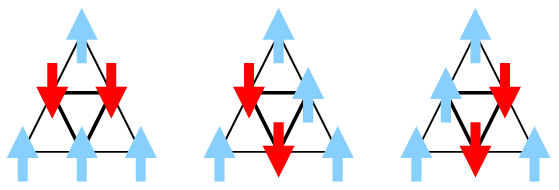}\label{fig:lattice}}}
\subfigure[Two of three $b$-spins pointing up]
{\resizebox*{0.9\columnwidth}{!}{\includegraphics{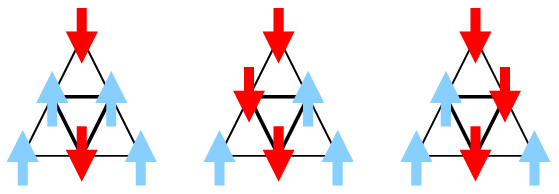}\label{fig:bond}}}
\par}
\caption{(Color online). The ln72 phase.
Given a configuration of the three $b$-spins (on the outer triangle), it can be shown, by enumerating all possibilities, that there are exactly three states of the $a$-spins (on the inner triangle).  The figure illustrates this for two configurations of $b$-spins; results for the other configurations can be seen from the Ising symmetry and the local rotational symmetry
of the triangular plaquettes. 
\label{zeroTzeroh}}
\end{figure}

The degeneracy of the ground state manifold can be understood 
by considering the energetics of a single large plaquette, {\em i.e.}, 
an $a$-spin trimer along with its enclosing $b$-spin trimer.
Representative plaquette configurations within the ground state are shown in Fig.~\ref{zeroTzeroh}.
We enumerate all possible plaquette energies in Table~\ref{energytable}.
As can be seen from the table, for any of the $2^3=8$ possible  configurations of the $b$-trimer,
there are three and only three configurations of the enclosed $a$-trimer
which are all within the ground state.\cite{fnote}   This means that the $b$-spins are effectively
free within the ground state manifold.  

In order to count the ground state degeneracy, we now turn to the unit cell.
Since the $b$-spins are free, and there are $3$ $b$-spins per unit cell,
this contributes $2^3 = 8$ configurations per unit cell to the ground state manifold. 
For any given configuration of the $b$-spins, each $a$-trimer in the lattice has a $3$-fold degeneracy.
Since there are two $a$-trimers per unit cell, these contribute a factor of  $3^2 = 9$ 
to the ground state degeneracy. 
The total degeneracy per unit cell in the ground state is therefore $8\times 9 = 72$,
as we showed in Sec.~\ref{s:entropy}.

The fact that the $b$-spins are effectively independent also means that the correlation function is  ``perfectly localized'': it is exactly zero beyond a distance $r_{bb}$ (the distance between two $b$-spins).  The correlation length $\xi$ is thus zero (where $\xi$ is defined as the asymptotic decay length of the correlation function at \emph{large} distances).

For comparison, at $T=h=0$, the triangular Ising AF has power-law correlations.
The kagome Ising AF is more frustrated than the triangular lattice case, since its ground state 
has exponentially decaying correlations.
We have shown here that the ground state of the TKL Ising AF in the frustrated regime has perfectly localized correlations,
making this model even more frustrated than either the triangular or kagome cases. 

\begin{table}[htb]
\[
\begin{array}{cc}
  & \sigma _a \\
 \sigma _b & 
\begin{array}{c|cccccccc}
  & \uparrow \uparrow \uparrow  & \downarrow \uparrow \uparrow  & \uparrow \downarrow \uparrow  & \uparrow \uparrow \downarrow  & \uparrow \downarrow
\downarrow  & \downarrow \uparrow \downarrow  & \downarrow \downarrow \uparrow  & \downarrow \downarrow \downarrow  \\
 \hline
 \uparrow \uparrow \uparrow  & \text{ 15} & \text{ -1} & \text{ -1} & \text{ -1} & \textbf{\blue{ -5}} & \textbf{\blue{ -5}} & \textbf{\blue{ -5}} & \text{  3} \\
 \downarrow \uparrow \uparrow  & \text{ 11} & \textbf{\blue{ -5}} & \text{ -1} & \text{ -1} & \text{ -1} & \textbf{\blue{ -5}} & \textbf{\blue{ -5}} & \text{  7} \\
 \uparrow \downarrow \uparrow  & \text{ 11} & \text{ -1} & \textbf{\blue{ -5}} & \text{ -1} & \textbf{\blue{ -5}} & \text{ -1} & \textbf{\blue{ -5}} & \text{  7} \\
 \uparrow \uparrow \downarrow  & \text{ 11} & \text{ -1} & \text{ -1} & \textbf{\blue{ -5}} & \textbf{\blue{ -5}} & \textbf{\blue{ -5}} & \text{ -1} & \text{  7} \\
 \uparrow \downarrow \downarrow  & \text{  7} & \text{ -1} & \textbf{\blue{ -5}} & \textbf{\blue{ -5}} & \textbf{\blue{ -5}} & \text{ -1} & \text{ -1} & \text{ 11} \\
 \downarrow \uparrow \downarrow  & \text{  7} & \textbf{\blue{ -5}} & \text{ -1} & \textbf{\blue{ -5}} & \text{ -1} & \textbf{\blue{ -5}} & \text{ -1} & \text{ 11} \\
 \downarrow \downarrow \uparrow  & \text{  7} & \textbf{\blue{ -5}} & \textbf{\blue{ -5}} & \text{ -1} & \text{ -1} & \text{ -1} & \textbf{\blue{ -5}} & \text{ 11} \\
 \downarrow \downarrow \downarrow  & \text{  3} & \textbf{\blue{ -5}} & \textbf{\blue{ -5}} & \textbf{\blue{ -5}} & \text{ -1} & \text{ -1} & \text{ -1} & \text{ 15}
\end{array}
\end{array}
\]
\caption{
Energy of a large triangle (consisting of 3 $b$-spins and 3 $a$-spins) as a function of the configurations of $a$ and $b$ spins.  For clarity of presentation, we present the results for the case $J_{aa}=-3$, $J_{ab}=-1$, $h=0$, but the form of the table is representative of the entire line $J_{aa}/|J_{ab}| < -1$.  The boldfaced numbers indicate the lowest-energy configurations, which have energy $J_{aa}+2J_{ab}$.  Note that each row contains exactly three boldfaced numbers.  See text for discussion.
\label{energytable}}
\end{table}

\subsection{Saturated ferromagnetic phase (phase I)}

At very high fields, when $h/|J_{ab}| > \max(4, 2|J_{aa}|/|J_{ab}|+2)$, we find that there is a unique ground state where all the $b$-spins and $a$-spins are up.  This state is easily seen to have magnetization $m=9$, entropy $s=0$, and energy $u= - 3h + 6\abs{J_{aa}} + 12\abs{J_{ab}} $ per unit cell.

        
\subsection{Ferrimagnetic phase (phase II)}
At lower fields $0 < h/|J_{ab}| < 4$ and when $J_{aa}/|J_{ab}| > -1$, there is a ferrimagnetic phase: the $a$-spins, being more numerous, align parallel to the field (up); the $b$-spins are then induced to point down due to the antiferromagnetic $J_{ab}$ interaction.  This phase has 
$m=3$, $s=0$, and $u= - 3h + 6\abs{J_{aa}} - 12\abs{J_{ab}}$ per unit cell.


\subsection{Log 9 phase (phase III)}
\begin{figure}[htb]
\centering
{\resizebox*{0.47\textwidth}{!}{\includegraphics{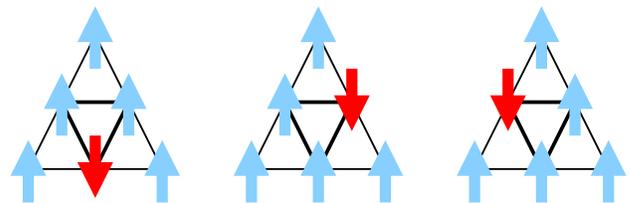}}}
\caption{(Color online). Phase III of the TKL Ising model.  When the field is quite strong, the $b$-spins (outer triangle) are all polarized, and each $a$-trimer (inner triangle) has three degenerate ground states.
\label{threestatepotts}}
\end{figure}

If $4 < h/|J_{ab}| < 2|J_{aa}|/|J_{ab}|+2$, the $b$-spins are completely polarized, but each $a$-triangle has 3 degenerate states (see Fig.~\ref{threestatepotts}).
Therefore, in this phase the system is equivalent to a set of non-interacting three-state Potts spins.  
This phase has $m=5$, $s=\ln 9=2.1972\dots$, and $u= - 5h - 2\abs{J_{aa}} + 4\abs{J_{ab}}$ per unit cell.  Again, the correlation function is perfectly localized, and the correlation length is $\xi=0$.

\subsection{Dimer phase (phase IV)}
\begin{figure}[htb]
\centering
{\resizebox*{0.47\textwidth}{!}{\includegraphics{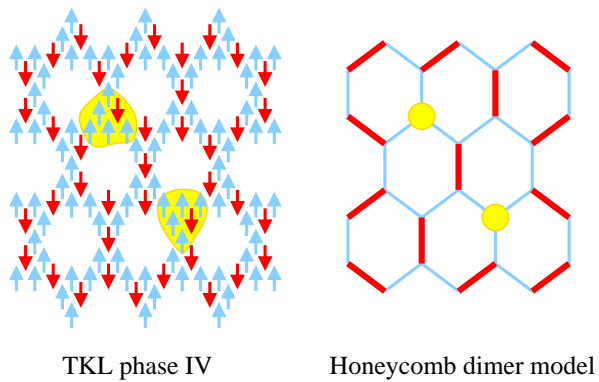}}}
\caption{(Color online). Phase IV of the TKL Ising model: Mapping of the spin configurations to configurations of dimers on a honeycomb lattice.  The yellow patches represent defects in the spin configuration, which correspond to vacancies (or monomers) in the dimer picture.
\label{honeydimer}}
\end{figure}

The most interesting situation occurs when $0 < h/|J_{ab}| < 4$ and $J_{aa}/|J_{ab}| < -1$.  
Table~\ref{energytable}  shows that  
the system will have the lowest energy if each
$b$-trimer has exactly one $b$-spin pointing down.  
Counting the number of ways to satisfy this constraint globally is a non-trivial problem.  The situation is the same as that for a kagome Ising AF at $T=0$ and $0<h<4|J_\text{kagome}|$, which has been studied before\cite{moessner2000,moessner2001,udagawa2002,moessner2003}.
The down-$b$-spins behave like a lattice gas on the kagome lattice with nearest-neighbor exclusion, at maximal density.
The ground states can be mapped to configurations of dimers occupying a honeycomb lattice (see Fig.~\ref{honeydimer}).
This problem has been solved exactly using the Pfaffian method\cite{kasteleyn1963}; the entropy per unit cell is 
\begin{align}
& \frac{1}{8\pi^2} \int_0^{2\pi} dp \int_0^{2\pi} dq ~ \ln \left(
        1 - 4\cos p \cos q + 4\cos^2 q
\right)
\nonumber\\
&= 0.3231\dots .
\end{align}
Therefore, the entropy of the TKL phase IV is also $0.3231\dots$ per unit cell, or $0.03590\dots$ per site.

The correlation function of the $b$-spins, $C_{bb} (\mathbf{r}) = \mean{\sigma_b(\mathbf{0})\sigma_b(\mathbf{r})}$, is equivalent to the dimer-dimer correlation function.  This correlation function decays as a power law, $1/r^2$, and nice visualizations in real space and reciprocal space are given in Ref.~\onlinecite{moessner2003}.  Thus, the model is in a critical phase, in the ``Villain-Stephenson universality class''.  The magnetization per unit cell is $m=3$ and the energy per unit cell is $u= - 3h - 2\abs{J_{aa}} - 4\abs{J_{ab}}$.

\subsection{Phase diagram}

\begin{figure}[!t]
\psfrag{h}{\LARGE $\frac{h}{\abs{J_{ab}}}$}
\psfrag{Jaa}{\Large $J_{aa}/\abs{J_{ab}}$}
\psfrag{Hc}{$h=2\abs{J_{aa}}+2\abs{J_{ab}}$}
\psfrag{+Sat}{$\begin{array}{l}         m=9 \\ s=0 \\
              \end{array}$}
\psfrag{+Ferri}{$\begin{array}{l}   \text{Ferrimagnetic} \\  m=3 \\ s=0 \\
            \end{array}$}
\psfrag{+Log9}{$\begin{array}{l}  
                                \\  m=5 \\          s=\ln 9 =2.1972 \\ 
      \end{array}$}
\psfrag{+Dimer}{$\begin{array}{l}  \text{Critical dimer phase} \\   m=3 \\ s=0.3231 
        \end{array}$}
\psfrag{ln72}{$\begin{array}{l}   \displaystyle\mathbf{V} \\ m=0 \\ s=\ln 72=4.2767      \end{array}$}
\psfrag{LRO}{$\begin{array}{l}   \displaystyle\mathbf{VI}    \end{array}$}
\centering
{\resizebox*{0.47\textwidth}{!}{\includegraphics{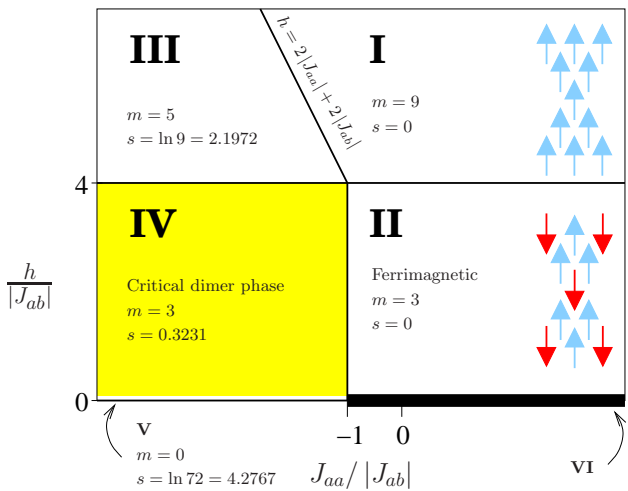}}}
\caption{
(Color online). Phase diagram of the TKL Ising model in the $(J_{aa},h)$ plane, for
  $J_{ab}<0$ (antiferromagnetic intertrimer coupling) and $T=0$.  The phase diagram is symmetric under a change of sign
of $h$.
The phases I', II', III', IV' are just mirror images of I, II, III, IV obtained by swapping up and down spins.  The yellow region represents the critical phases with power-law correlations; note that it does not include the thin line at $h=0$, which is phase V, the $\ln 72$ phase described in the text.  The thick line (VI) persists as a true ferrimagnetic phase transition at finite $T$; all other lines turn into crossovers.
\label{phase_diag_Jab-1_T0}}
\end{figure}

\begin{figure}[htb]
\psfrag{h}{\LARGE $\frac{h}{\abs{J_{ab}}}$}
\psfrag{Jaa}{\Large $J_{aa}/\abs{J_{ab}}$}
\psfrag{Hc}{$h=2\abs{J_{aa}}-2\abs{J_{ab}}$}
\psfrag{ferro}{$\begin{array}{l}   \text{Ferrimagnetic} \\  m=3 \\ s=0 \\
            \end{array}$}
\psfrag{ln9}{$\begin{array}{l}  
                                \\  m=5 \\          s=\ln 9 =2.1972 \\ 
      \end{array}$}
\psfrag{ln72}{$\begin{array}{l}   \displaystyle\mathbf{V} \\ m=0 \\ s=\ln 72=4.2767      \end{array}$}
\psfrag{LRO}{$\begin{array}{l}   \displaystyle\mathbf{VI} \end{array}$}
\centering
{\resizebox*{0.47\textwidth}{!}{\includegraphics{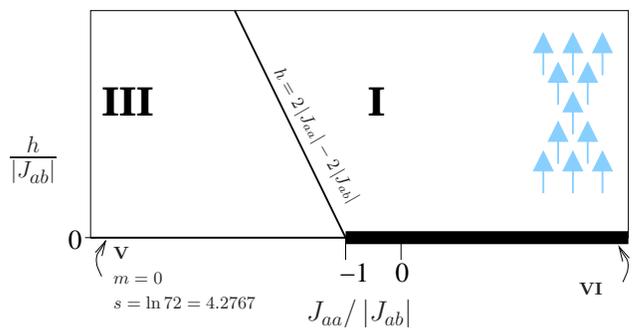}}}
\caption{(Color online). Phase diagram of the TKL Ising model in the $(J_{aa},h)$ plane, for $J_{ab}>0$ (ferromagnetic intertrimer coupling) and $T=0$.
\label{phase_diag_Jab1_T0}}
\end{figure}

We now combine the above results in order to report the full
phase diagram of the TKL Ising model.
Fig.~\ref{phase_diag_Jab-1_T0} shows the phase diagram for 
antiferromagnetic intertrimer coupling, $J_{ab}<0$,
and in Fig.~\ref{phase_diag_Jab1_T0} we show the phase diagram for 
ferromagnetic intertrimer coupling, $J_{ab}>0$.  
In both cases, the phase diagram is symmetric under 
for $h\rightarrow -h$, with simultaneous change of sign of all spins.
The entropy and magnetization change discontinuously across every
zero-temperature phase boundary.  Exactly on the phase boundaries and
intersections of these boundaries, the entropy will be higher than in either adjacent
phase, because the system can choose from states within each set of ground
states.  

Note that the high-field phases (I and III) are common to both phase diagrams.
The more interesting case is that of 
Fig.~\ref{phase_diag_Jab-1_T0}, 
which has antiferromagnetic
intertrimer coupling $J_{ab}<0$, and more phases at intermediate field strength. 
Right at $h=0$ in both phase diagrams, the ground state is phase V, the spin liquid with
residual entropy $s_0= {\rm ln 72}$ that we discussed in Sec.~\ref{sec:zerofield}.  
When $J_{ab}<0$, the application of an infinitesimal field 
induces a critical state with power law correlation functions (phase IV), 
which we have mapped to the problem of hard-core dimers on a honeycomb lattice.



\section{Finite temperature and finite field\label{finiteTfiniteh}}

Most of the phase transitions in the zero-temperature phase diagrams
are destroyed by thermal fluctuations.  
The clear exception is phase VI, which has spontaneously broken
$Z_2$ symmetry.  This long-range ordered phase survives at finite temperature,
and has a true phase transition at a Curie temperature $T_c\ne0$.
Since $h$ is a relevant perturbation, this finite-temperature phase transition 
is destroyed at any finite field, leaving only a crossover.  


The other transitions in 
Figs.~\ref{phase_diag_Jab-1_T0}  and \ref{phase_diag_Jab1_T0}
are not characterized by a \emph{spontaneously} broken symmetry with an order parameter.  
Therefore, they cannot persist at finite $T$ as traditional order-disorder transitions.
However, a more subtle analysis is required to understand whether the critical phase, phase IV, persists at finite $T$ (bounded, e.g., by a Kosterlitz-Thouless transition curve).
The situation is quite similar to that for the kagome Ising AF described in Ref.~\onlinecite{moessner2003}. 
Based on the table of ground state energies, Table~\ref{energytable},
it is possible to enumerate the types of defects that can occur in phase IV.
These defects correspond to breaking a dimer into two monomers (see Fig.~\ref{honeydimer}), and can only be created in pairs.  
Each defect has energy $\Delta U = \min(h,  4|J_{ab}|-h)$.  The entropy $\Delta S$ associated with creating a defect pair is at least $O(\ln L^2)$, because one may choose to break any of the $O(L^2)$ dimers; in fact the entropy is even greater than this because the resulting monomers can be moved apart, resulting in many new configurations.  Hence at any finite $T$, the density of monomers is finite.  A theorem of Lieb and Heilman\cite{heilman1972} states that monomer-dimer models cannot have phase transitions at any finite density of monomers.  Therefore, \emph{the critical phase only exists at $T=0$, and it is destroyed at finite $T$}.

However, if the density of monomers is low, the correlation length $\xi$ may still be long; na\"ively, one might expect $\xi^2 \sim 1/n_\text{monomers} \propto \exp (E_\text{monomer}/T)$, but a more careful treatment accounting for the effective Coulomb attraction between monomers gives the result $\xi^2 \sim 1/n_\text{monomers} \propto \exp (8E_\text{monomer}/7T)$.\cite{moessner2003}
This scenario is consistent with the size-dependent peaks in the susceptibility that we find from Monte Carlo simulations (see Sec.~\ref{montecarlo}).

\section{Monte Carlo Simulations \label{montecarlo}}

\begin{figure}[t]
{\centering
  \subfigure[~Heat capacity\label{fm:ct}]{\resizebox*{0.8\columnwidth}{!}{\includegraphics{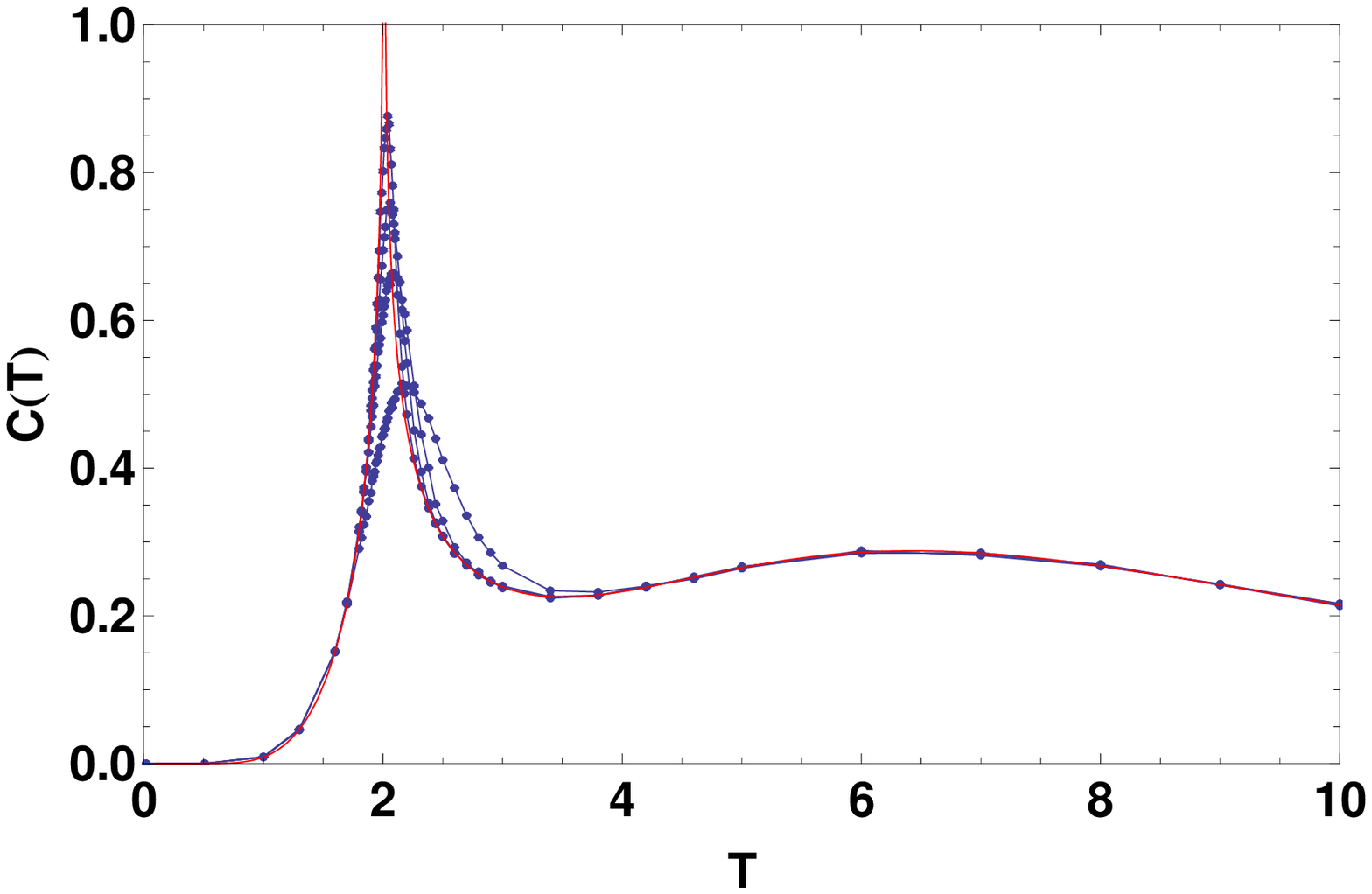}}} \par}
{\centering
  \subfigure[~Susceptibility\label{fm:xt}]{\resizebox*{0.8\columnwidth}{!}{\includegraphics{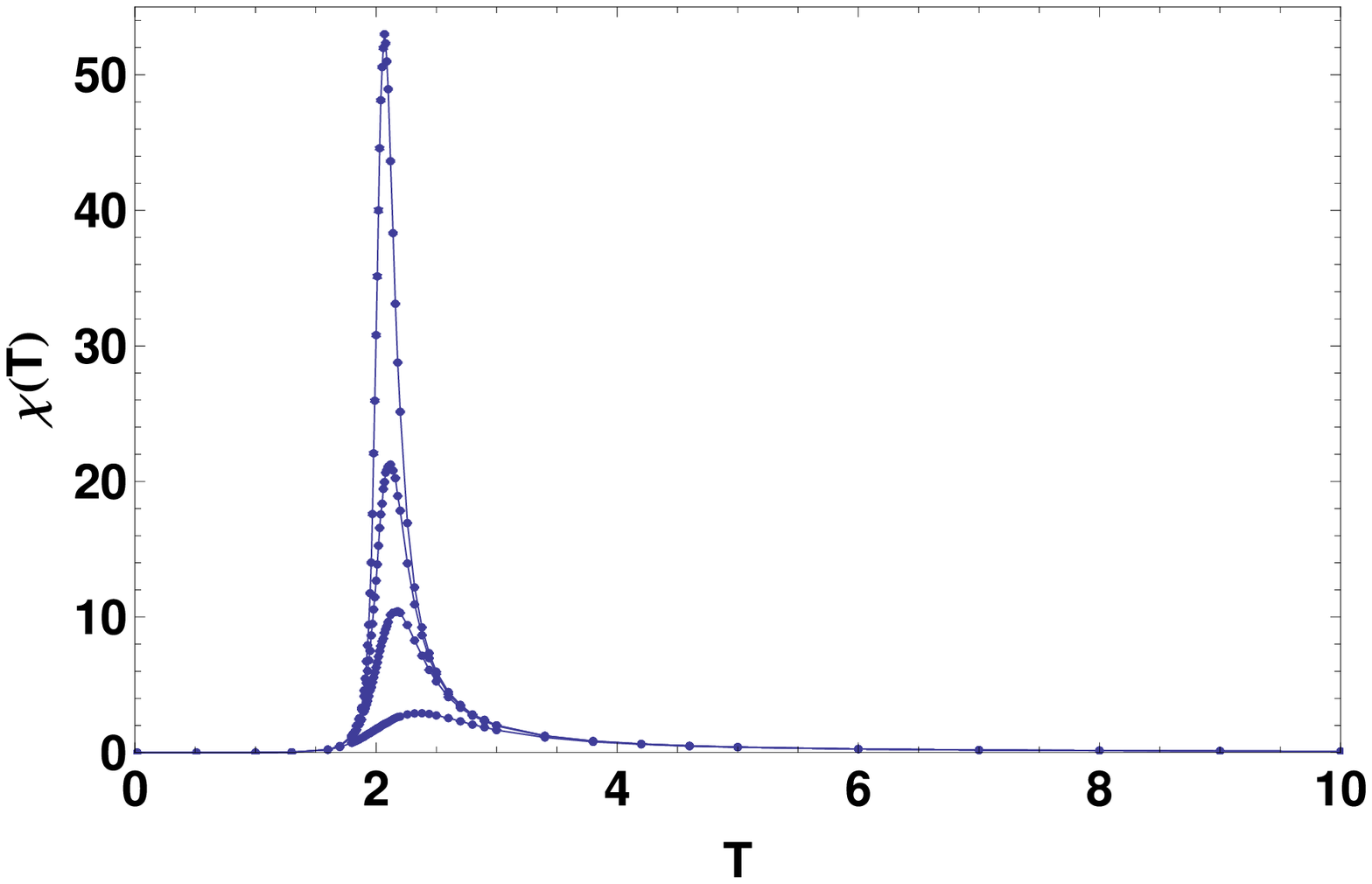}}}
   \par}
\caption{(Color online).Temperature dependence of heat capacity and
  susceptibility from Monte Carlo simulation for $J_{aa}=5$ and $J_{ab}=+1$ with system
  sizes $L=12,24,36,60$. Red line (no dots) represents the exact solution of specific
  heat. Peaks become taller and narrower as $L$ increases.}
\label{fm}
\end{figure}
\begin{figure}[t]
{\centering
  \subfigure[~Specific heat\label{af1:ct}]{\resizebox*{0.8 \columnwidth}{!}{\includegraphics{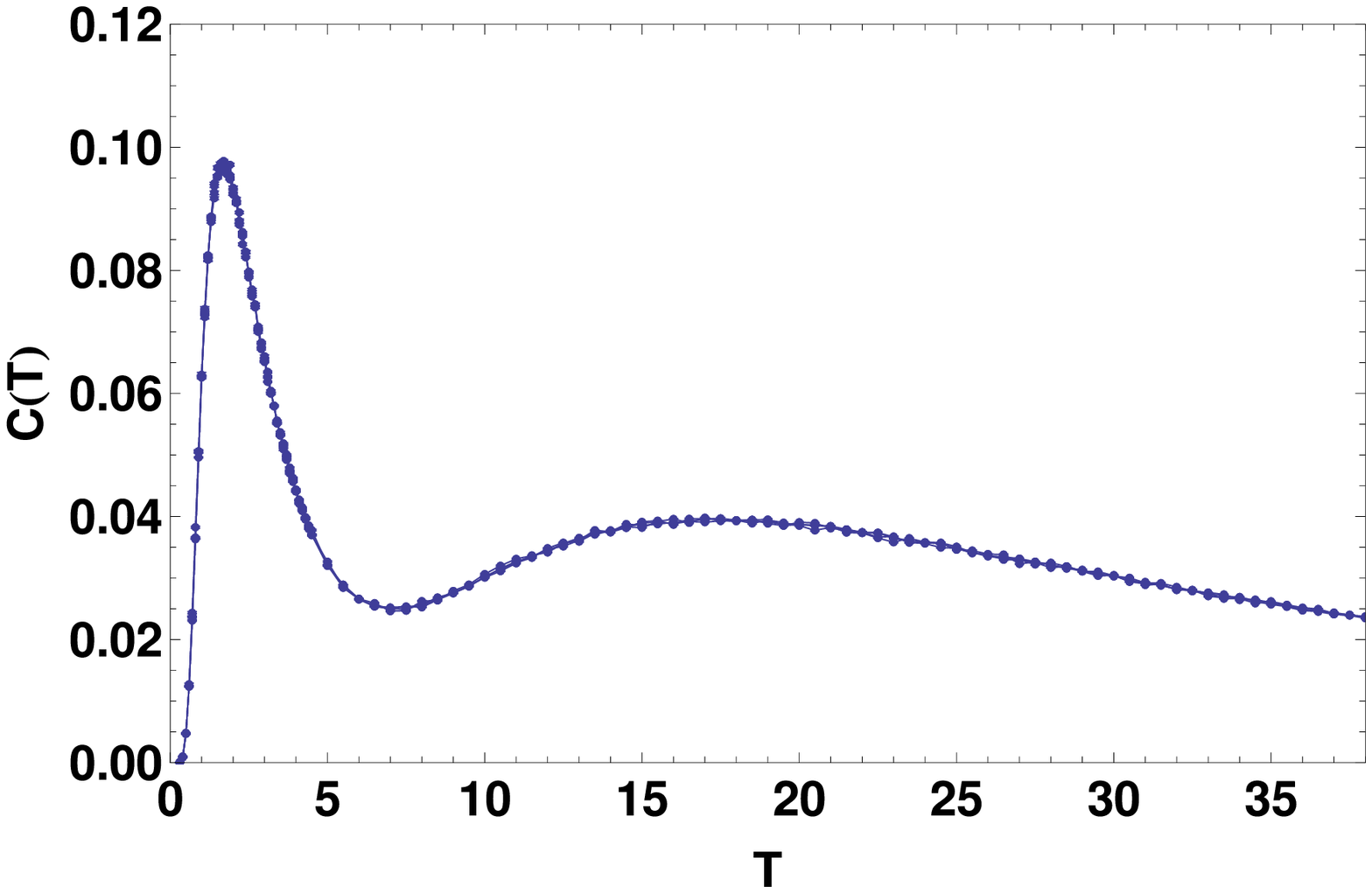}}} \par}
{\centering
  \subfigure[~Inverse susceptibility ($1/\chi(T)$)\label{af1:ix}]{\resizebox*{0.8\columnwidth}{!}{\includegraphics{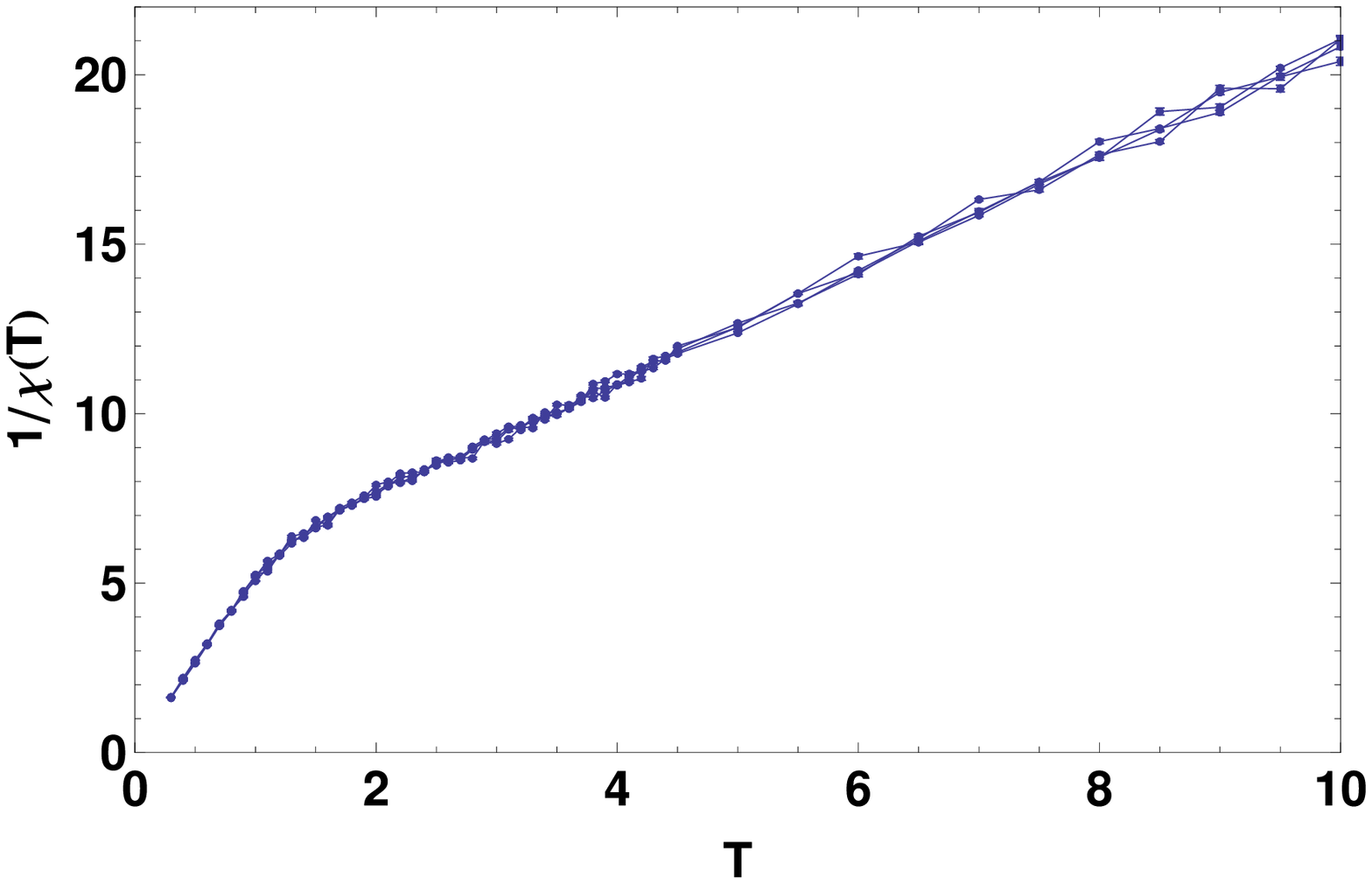}}}
   \par}
{\centering
  \subfigure[~$T\chi(T)$\label{af1:tx}]{\resizebox*{0.8\columnwidth}{!}{\includegraphics{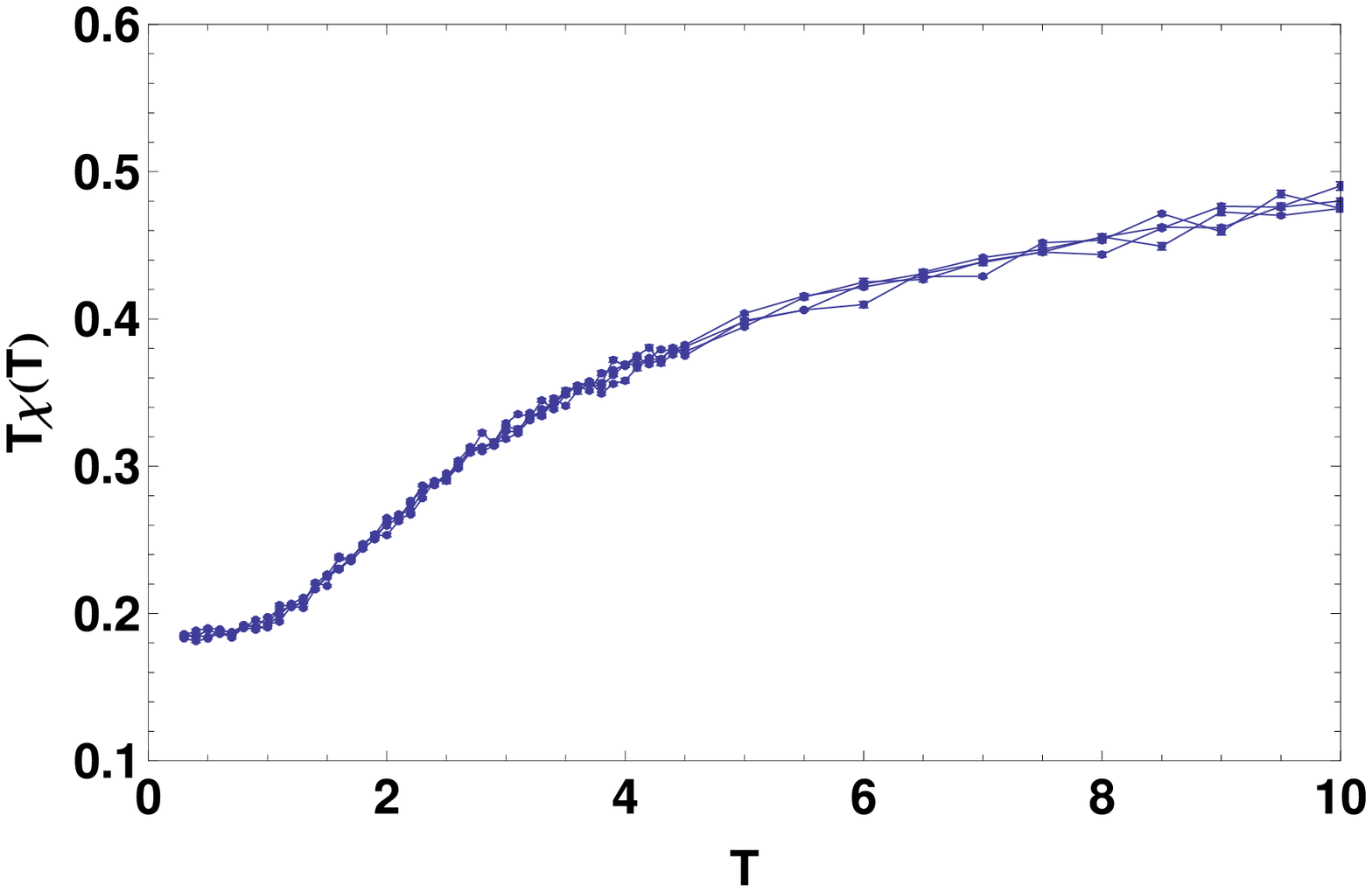}}}
   \par}
\caption{(Color online). Temperature dependence of heat capacity and inverse susceptibility ($1/\chi(T)$) and $T\chi(T)$ from Monte Carlo simulation for $J_{aa}=-10$ and $J_{ab}=-1$ with system sizes $L=12,24,36,60$.}
\label{af1}
\end{figure}
\begin{figure}[t]
{\centering
  \subfigure[~Specific heat\label{af2:ct}]{\resizebox*{0.8 \columnwidth}{!}{\includegraphics{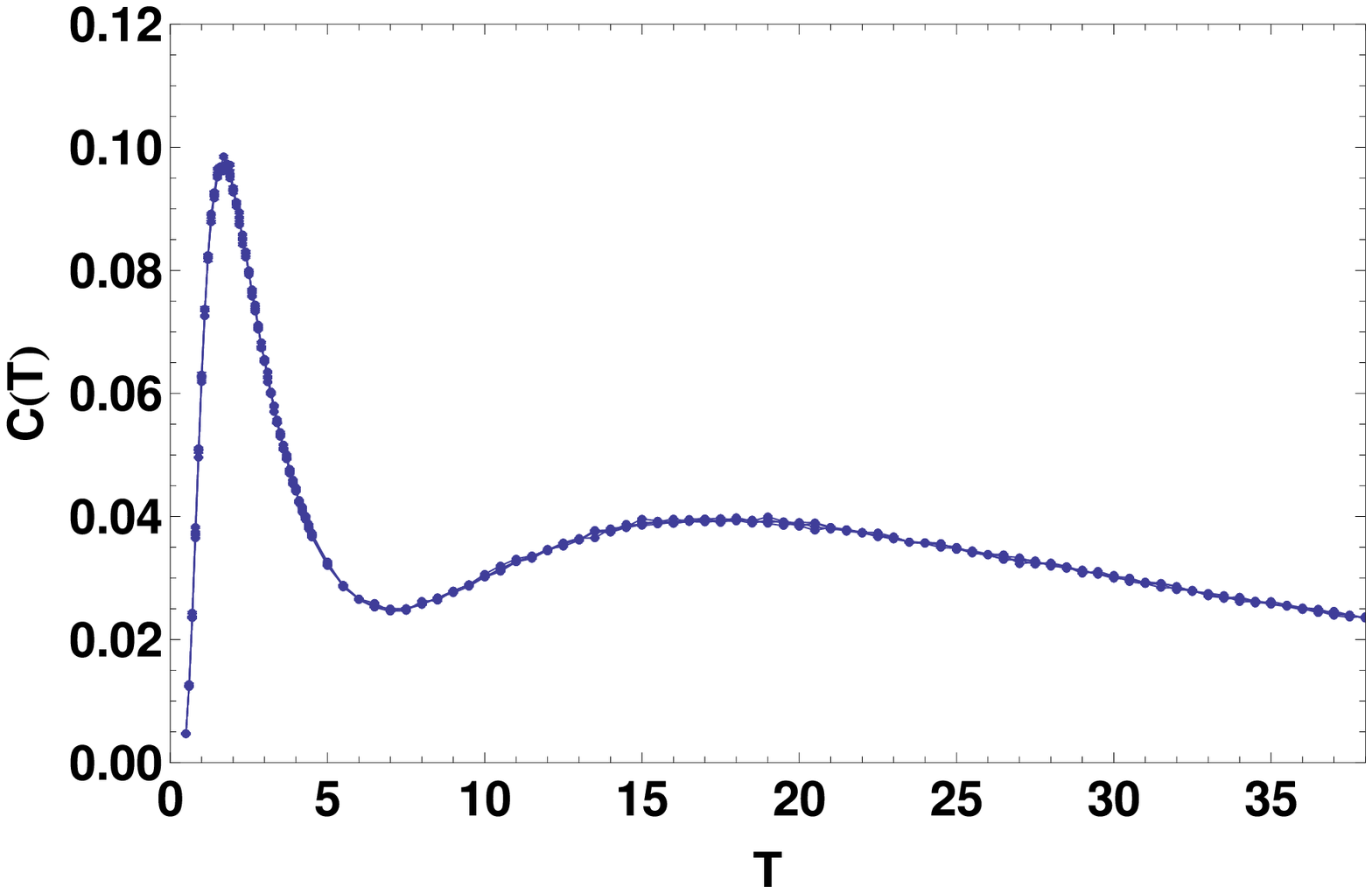}}} \par}
{\centering
  \subfigure[~Inverse susceptibility ($1/\chi(T)$) \label{af2:ix}]{\resizebox*{0.8\columnwidth}{!}{\includegraphics{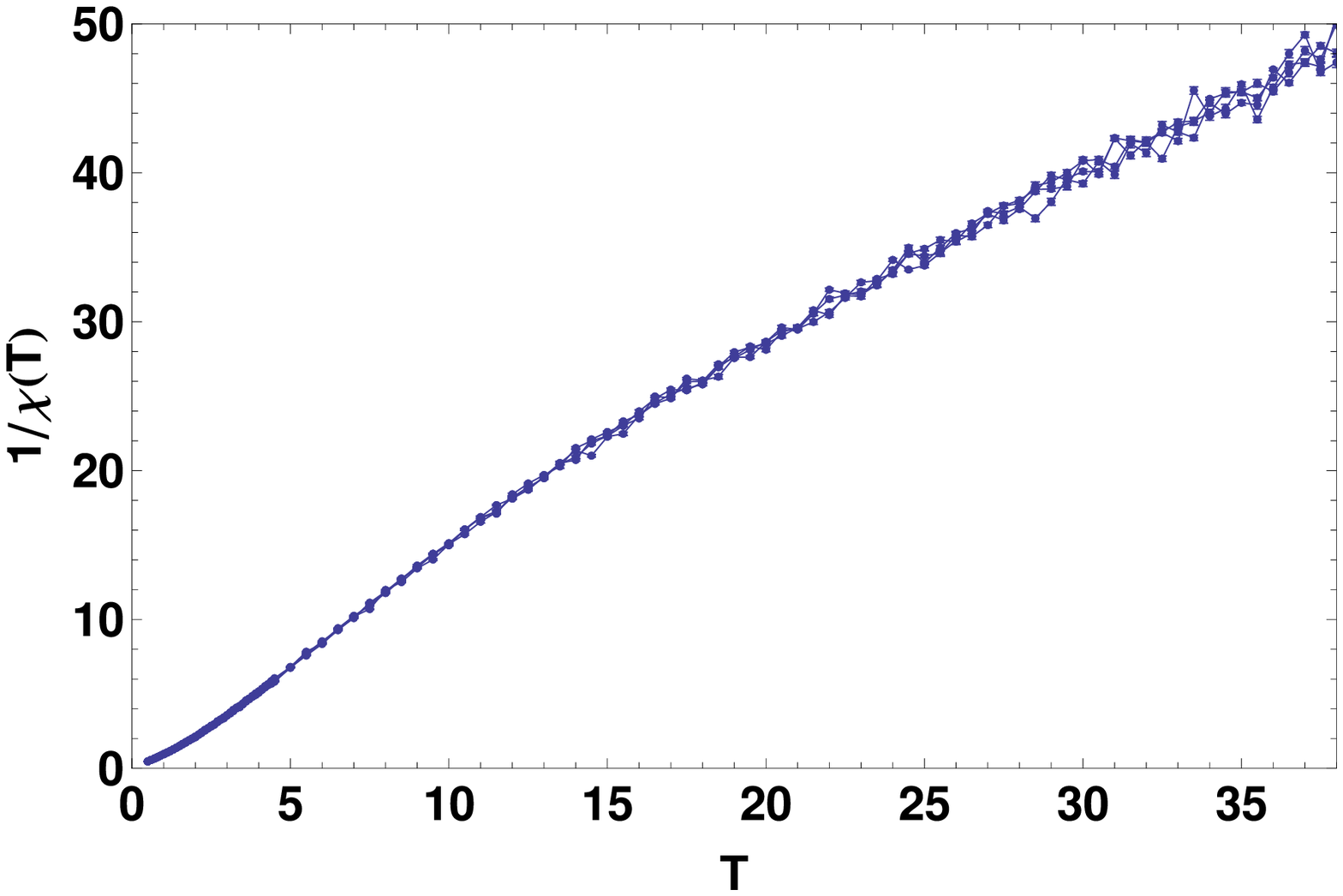}}}
   \par}
{\centering
  \subfigure[~$T\chi(T)$ \label{af2:tx}]{\resizebox*{0.8\columnwidth}{!}{\includegraphics{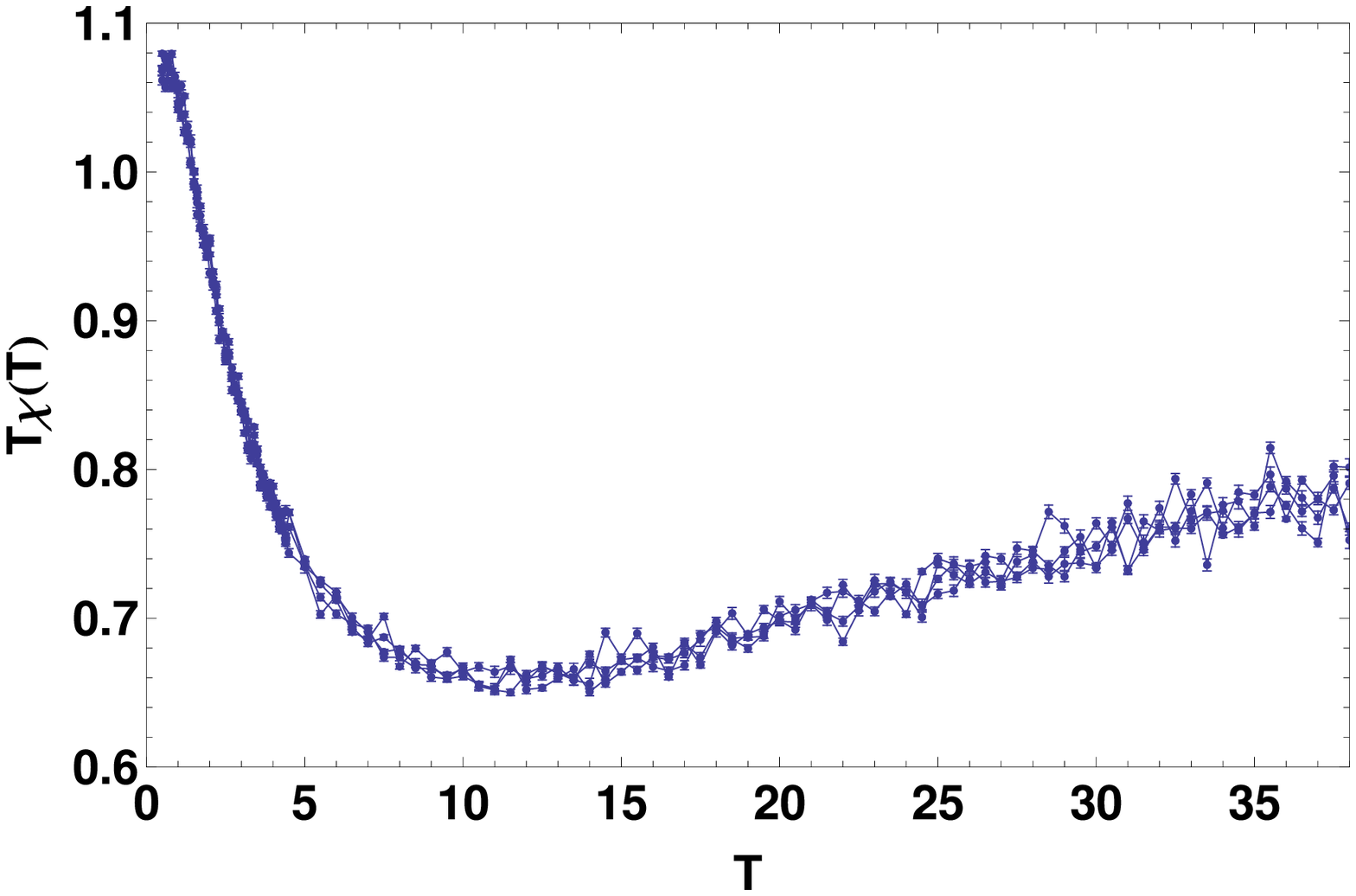}}}
   \par}
\caption{(Color online). Temperature dependence of heat capacity, inverse susceptibility ($1/\chi(T)$) and $T\chi(T)$ from Monte Carlo simulation for $J_{aa}=-10$ and $J_{ab}=1$ with system sizes $L=12,24,36,60$.}
\label{af2}
\end{figure}

In this section we present the results of Monte Carlo (MC) simulations of the TKL
Ising model, Eq.~(\ref{e:hamiltonian}), for various combinations of parameters.
The simulations corroborate our analytic predictions
and also allow us to perform calculations at finite $h$ and $T$
as well as to compute the magnetization and susceptibility.

We use the Wolff algorithm for $J_{aa}>0$ at $h=0$, and the Metropolis
algorithm for $J_{aa}<0$ at $h=0$ and $h \neq 0$.  The system sizes we use are
$L=12, 24, 36, 60$, where $L$ is the 
length of the underlying triangular
lattice, so that the total number of spins is $N = \frac{9}{16} L^2$; for
periodic boundary conditions, $L$ should be a multiple of 3 in order to avoid
introducing boundary defects and additional frustration.  

In order to evaluate the heat capacity $C$ and magnetic susceptibility $\chi$,
we use the fluctuation-dissipation theorem:
\begin{align}
        C &=\frac{\mean{H^2} - \mean{H}^2}{NT^2},\\
        \chi   &=\frac{\mean{M^2} - \mean{M}^2}{NT}, \label{chi}
\end{align}
where $\mean{H}$ and $\mean{M}$ are the Monte Carlo averages of the
total energy (i.e., the Hamiltonian) and magnetization, respectively.
We define the sublattice magnetizations as 
\begin{equation} 
 m_a= \frac{1}{N_a}\ \sum_{i \in a} \sigma_{ai}~,
\end{equation}
\begin{equation}
 m_b= \frac{1}{N_b}\ \sum_{i \in b} \sigma_{bi}~,
\end{equation}
where $N_a$ is the number of $a$-spin sites, 
and $N_b$ is the number of $b$-spin sites.


\subsection{Zero magnetic field}  

We first show Monte Carlo results at zero field. 
Fig.~\ref{fm} shows the 
temperature evolution of the heat capacity and susceptibility 
in a representative unfrustrated case, $J_{aa} = 5 J_{ab} >0$.
The Monte Carlo results for the heat capacity are consistent with the exact results in
Sec.~\ref{zerotemperature}; the peak in the heat capacity 
in our simulations becomes taller
and narrower as $L$ increases, tending towards the exact solution for
$L=\infty$.

Figs.~\ref{af1} and
~\ref{af2} show results for two frustrated parameter combinations
($J_{aa}/|J_{ab}| < -1$).  
In both cases, the susceptibility shows a marked  difference from the
case of ferromagnetic intratrimer coupling, $J_{aa}>0$.
For $J_{aa}>0$, the susceptibility shows a sharp peak at the Curie
temperature $T_c$, whereas for $J_{aa}<0$, 
it tends to $\infty$ as $T\rightarrow 0$.

Unlike the free energy, heat capacity, entropy, and internal energy,
the susceptibility depends on the sign of $J_{ab}$.
Our predictions about the
susceptibility can be used to distinguish whether a physical TKL Ising
system has ferromagnetic or antiferromagnetic coupling between the $a$
and $b$ sublattices.  For $J_{ab}>0$, the inverse susceptibility
$1/\chi(T)$ shows two linear pieces, with a crossover at the lower
coupling constant $J_{ab}$.  However, when $J_{ab}<0$, the inverse
susceptibility approaches zero much slower than the $J_{ab}>0$ case.

The difference between antiferromagnetic and ferromagnetic
coupling $J_{ab}$ is even more evident in the plots of $T \chi$.  At low temperatures,
$T\chi$ tends to a small finite constant as $T \rightarrow 0$ if the intertrimer coupling
is antiferromagnetic, $J_{ab}<0$; whereas, if the intertrimer coupling
is ferromagnetic, $J_{ab}>0$, $T\chi$ goes through a minimum  around $T\sim J_{aa}$, increases sharply, and saturates at a finite value as $T\rightarrow 0$.



\subsection{Finite magnetic field}    

\begin{figure}[!t]
\centering
\includegraphics[width=8cm,clip]{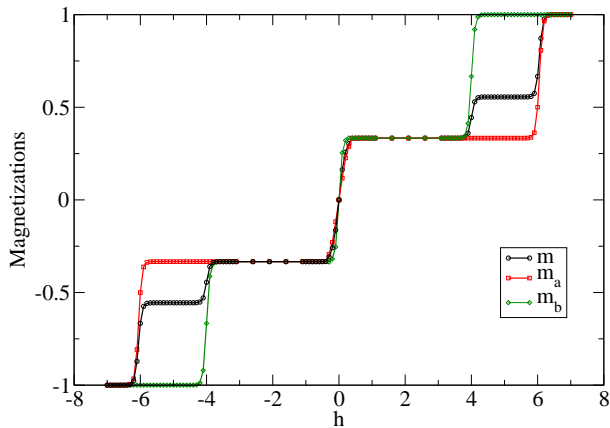}
\caption{(Color online). Magnetization $m$ vs. external field $h$ from Monte Carlo simulation for $J_{aa}=-2$, $J_{ab}=-1$ at
  $T=0.1$ for $L=12$.  Sublattice magnetizations $m_a$ and $m_b$ are also shown.}
\label{mh}
\end{figure}

\begin{figure}[!t]
\centering
\includegraphics[width=8cm,clip]{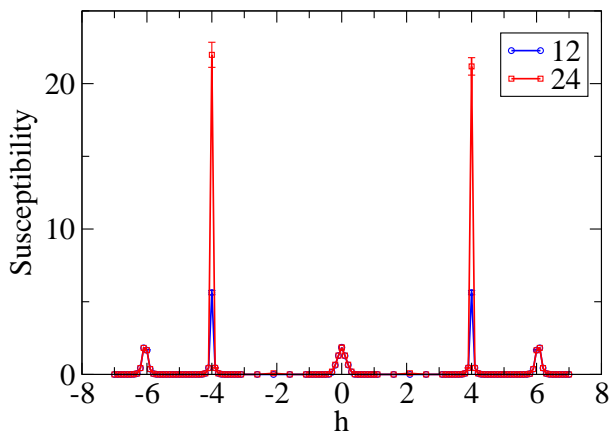}
\caption{\label{xh}
(Color online). Susceptibility $\chi$ vs. external field $h$  from Monte Carlo simulation for $J_{aa}=-2$, $J_{ab}=-1$ at
  $T=0.1$ for $L=12 $.                                 }
\end{figure}

In Figs.~\ref{mh} and \ref{xh}, we show Monte Carlo results at finite field $h$ and low temperature $T=0.1$ for $J_{aa}=-2$ and $J_{ab}=-1$.  
The magnetization curves in Fig.~\ref{mh} have a series of steps and plateaux (as is typical of frustrated spin systems).
Starting from $h=0$, as a field is applied, both the $a$ and $b$ sublattices immediately respond
as the critical dimer phase is induced,
developing (normalized) sublattice magnetizations  of $m_a=m_b=1/3$.
As field is increased, the $b$-spins are more easily polarized, while the $a$-sublattice spins  only become fully polarized when the magnetic field is strong enough.

The susceptibility as a function of applied field has a series of peaks, as shown in Fig.~\ref{xh}.   
The peaks at $h=\pm 4$ increase with increasing system size from $L=12$ to $L=24$.
This indicates that the correlation length (in the vicinity of the peaks) is comparable to, or larger than, the system size.  This does not indicate a true finite-temperature phase transition.  Rather, because phase IV is critical (with infinite correlation length), the correlation length diverges 
as temperature is lowered toward this phase.  At the low temperatures we have simulated,
the correlation length is comparable to our system sizes.  See Sec.~\ref{finiteTfiniteh}.


\section{Discussion \label{discussion}}

Phases I,II,and III can easily be simulated using the Metropolis algorithm.  However, phase IV has extremely slow dynamics when simulated using the Metropolis algorithm at low temperature -- this could be described as glassy dynamics.  
We believe that the simulations may be made more efficient by using geometric cluster methods\cite{heringa1998} or by augmenting the Metropolis algorithm with directed-loop updates tailored to the honeycomb dimer state.\cite{krauth-2003,sandvik-2006}  
The slow dynamics under the Metropolis algorithm may, however, be representative of the true dynamics in a physical realization of the TKL Ising antiferromagnet.  

It may be possible to stabilize more phases at finite temperature by introducing
an appropriate perturbation.  
For example, introducing coupling in the third direction may be sufficient to 
stabilize the critical phase it finite T.  
In addition,  it may be possible to induce a Kasteleyn transition in phase IV by applying some kind of orienting field ({\em e.g.}, uniaxial strain) to the TKL, similar to what was suggested by Moessner and Sondhi\cite{moessner2003} on the Kagome lattice.
In light of existing studies on triangular and kagome lattices, it
is reasonable to expect that adding next-nearest-neighbor interactions
to the TKL will produce an even richer phase diagram, like Kagome lattice.~\cite{takagi93}


 
\subsection{Comparison with Other Frustrated Ising Models}

In Table~\ref{comparison-table}, we present a comparison of the
spin-spin correlations and the residual entropy among the frustrated
TKL, triangular, and kagome lattices.  The simplest example of a
geometrically frustrated system is the triangular Ising
antiferromagnet (TIAF), {i.e.}, a set of Ising spins on a triangular
lattice with antiferromagnetic pairwise couplings.  Due to the
presence of odd cycles in the lattice graph, it is impossible for all
pairs of nearest neighbor spins to be simultaneously antiparallel.  As
a result, the antiferromagnetic interactions are unable to produce
long-range order even at zero temperature.  Instead, at zero
temperature, the TIAF has a quasi-long-range-ordered state in which
the correlations decay with distance as $r^{-1/2}$.
\cite{stephenson64,stephenson70} This ground state is macroscopically
degenerate, with a zero-temperature residual entropy of $0.3231k_B$ per spin. \cite{wannier50,syozi50a,syozi50b,foot1}  This number is the same as the entropy per unit cell of the random dimer model on a honeycomb lattice, and it crops up in many other places.
Although the TIAF has no phase transition in zero field, the
application of a finite field produces a surprisingly rich phase
diagram.  A small field induces a Berezinskii-Kosterlitz-Thouless
(BKT) transition to a `spin crystal' phase that breaks translational
symmetry and has long-range order -- the correlation function
oscillates with distance but does not decay.  At larger fields the
crystalline order is destroyed via a transition in the 3-state Potts
universality class.  In this limit the TIAF is related to Baxter's
exactly soluble hard hexagon lattice gas model.\cite{baxter-1980}

Another frustrated spin system is the antiferromagnetic Ising model on
a kagome lattice, formed by periodic removal of a quarter of the sites
from the triangular lattice.  Unlike the TIAF, the KIAF in the absence
of field has pair correlations that decay
exponentially at all temperatures including
$T=0$. \cite{syozi,suto1981} Its ground state entropy is $1.5055k_B$ per unit cell or $0.5018 k_B$
per spin,\cite{kano53} higher than that of the TIAF because the
smaller coordination number allows more freedom.  At finite $h$, there is a different spin
liquid state that can be mapped to random dimers on a honeycomb
lattice.  In this state the spin-spin correlation function decays as a
power law, $1/r^2$.  The residual entropy is $0.3231k_B$ per unit cell, or $0.1077k_B$ per spin.

The ground state of the TKL Ising AF without applied field is
even more frustrated than that of the kagome lattice:
the correlation function becomes exactly zero beyond a certain cutoff radius.  
The residual entropy per spin is $\frac{1}{9}\ln 72 = 0.4752\dots$.  
At finite $h$, there is a correlated, critical spin liquid state
which we have mapped to hard core dimers on a honeycomb lattice.
This state has a residual entropy of $0.3231k_B$ per unit cell, or $0.03590$ per spin.
(See Table~\ref{comparison-table}.)

\begin{table}
\begin{ruledtabular}
\begin{tabular}{ccc}
Lattice              &  Entropy &  Spin-spin correlation    \\ \hline
Triangular           &  $0.3231\dots$  &  $r^{-1/2}$  \\
Triangular in field  &  $0$            &  long-range-ordered  \\
Kagome               &  $1.5055\dots$ &  $e^{-r/\xi}$    
\\
Kagome in weak field &  $0.3231\dots$ &  $r^{-2}$ \\
TKL                  &  $\ln 72$       &  $0$ for $r\geq r_{bb}$ \\
TKL in weak field    &  $0.3231\dots$  &  $r^{-2}$  \\ 
TKL in medium field  &  $\ln 9$        &  $0$ for $r\geq r_{bb}$ 
\end{tabular}
\end{ruledtabular}
\caption{Comparison of various frustrated Ising models at $T=0$.  Residual entropies are quoted per unit cell; they are the logarithms of irrational numbers, unless otherwise stated.
\label{comparison-table}}
\end{table}

\subsection{Comparison with Experiment}

As discussed in the introduction, the recently fabricated family of compounds
$\mbox{Cu}_{9}\mbox{X}_2(\mbox{cpa})_{6}\cdot x\mbox{H}_2\mbox{O}$
have an arrangement of the copper sites which forms a triangular kagome lattice.  
There is no evidence of 
spontaneous magnetization down to
at least $T=1.7$K,\cite{maruti94} consistent with a spin liquid ground state, which
indicates that $J_{aa}$ is antiferromagnetic.
However, there is no agreement yet as to whether $J_{ab}$ is
ferromagnetic or antiferromagnetic.\cite{maruti94,strecka07}

In studying the magnetic susceptibility, several groups find that the
slope of $1/\chi$ versus $T$ is roughly linear at high
temperature, but as $T$ is lowered, the slope increases.\cite{maruti94,haar95,mekata98}.
This is consistent with either $J_{ab}$ ferromagnetic or antiferromagnetic,
as seen in Figs.~\ref{af1:ix} and~\ref{af2:ix}.  
A distinction can be made, though, by the behavior of $T \chi$.
Whereas $T \chi$ saturates to a finite value at low $T$ for antiferromagnetic
intertrimer coupling $J_{ab}<0$, $T \chi$ reaches a minimum at intermediate $T$
before saturating at a finite value as $T \rightarrow 0$ if   
the intertrimer coupling is {\em ferromagnetic}, 
$J_{ab}>0.$
To the extent that the 
$\mbox{Cu}_{9}\mbox{X}_2(\mbox{cpa})_{6}\cdot x\mbox{H}_2\mbox{O}$
materials can be described by an Ising TKL model like the one in this paper,
our calculations indicate that the intertrimer coupling
must be ferromagnetic, $J_{ab}>0$,
so that upon application of a field, the system should be in 
Phase III, rather than Phase IV.
Since there is a striking difference in the residual entropies
in these two phases ($s_0 = \ln 9=2.1972\cdots$ in Phase III
{\em vs.} $s_0 = 0.3231\cdots$ in Phase IV),
heat capacity measurements will also be useful in distinguishing these phases.
Other future experiments, including neutron scattering, NMR, and $\mu$SR, can also provide useful data for comparing to models of frustrated
magnetism on the TKL. 



\section{Conclusions \label{conclusions}}

In conclusion, we have studied an Ising model on the triangular kagome
lattice (TKL) with two interactions $J_{aa}$ and $J_{ab}$, temperature
$T$, and external field $h$.  We have calculated the complete phase
diagram in $(J_{aa}, J_{ab}, h, T)$ parameter space \emph{exactly}.
Furthermore, we have obtained exact results for thermodynamic
quantities (free energy, energy, heat capacity, and entropy) at all
$T$ for $h=0$, and at all $h$ for $T=0$, and plotted them for
representative cases.

In the experimentally relevant regime, $\abs{J_{aa}} \gg
\abs{J_{ab}}$, if $J_{aa}$ is ferromagnetic, the specific heat shows a
broad hump corresponding to intra-trimer ordering, as well as a sharp
peak at lower temperatures due to the onset of true long-range order.
If $J_{aa}$ is antiferromagnetic there are two broad features.

We have computed the magnetization $M(T,h)$ and susceptibility $\chi(T,h)$ in various regimes using Monte Carlo simulations.  
To the extent that experiments on the $\mbox{Cu}_{9}\mbox{X}_2(\mbox{cpa})_{6}\cdot x\mbox{H}_2\mbox{O}$
materials can be compared with an Ising TKL model, our calculations
indicate that $J_{ab}>0$.

We find that at strong frustration and zero field, as temperature is
reduced, the model enters a spin liquid phase with residual entropy $s_0 = \ln 72$ per unit cell, with ``perfectly localized'' correlations.  This stands in contrast with the triangular
and kagome Ising antiferromagnets, whose residual entropies cannot be expressed in closed form.  

The most interesting feature of the model is a correlated critical
spin liquid phase (with power-law correlations) that appears at strong
frustration, weak fields, and zero temperature.
We have mapped this phase to hard core dimer coverings of a honeycomb
lattice.  The critical power-law correlations are reduced to
exponential correlations at finite $T$, but the correlation length may
still be large.
Such a phenomenon would be detectable by neutron scattering measurements.

\acknowledgments
It is a pleasure to thank T.~Takagi, M.~Mekata, G.~Ortiz, N.~Sandler, and M.~Ma for helpful discussions.
This work was supported by Purdue University and Research Corporation (Y.~L.~L. and D.~X.~Y.). 
E.~W.~C. is a Cottrell Scholar of Research Corporation.

\appendix
\section{Mean-Field Approximations \label{meanfield}}

For pedagogical purposes, we examine the zero-field TKL Ising model using two mean-field approaches.  Compared with the exact solution (see Fig.~\ref{zerofieldphasediagram}), we see that such approaches can be quite misleading in the frustrated regime $J_{aa}/|J_{ab}|<-1$.

In the simplest mean-field approximation (MF1), every spin is assumed to fluctuate thermally in a mean field determined by the average magnetizations of its neighbors.  This leads to a pair of simultaneous equations for the spontaneous magnetizations of the $a$- and $b$-sublattices, $m_a=\mean{\sigma_a}$ and $m_b=\mean{\sigma_b}$,
 \begin{align}
 m_a &= \tanh \beta(2J_{aa}m_a+2J_{ab}m_b),
\nonumber\\
 m_b &= \tanh \beta(4J_{ab}m_a).
 \label{mft}
 \end{align}
where $\beta=1/T$.
Linearizing the tanh function in $\mean{\sigma_a}$ and $\mean{\sigma_b}$ gives the critical temperature,
  \begin{align}
        T_c^\text{MF1} = J_{aa} + \sqrt {J_{aa} {}^2 + 8 J_{ab} {}^2}.
  \end{align}

We also present a more advanced mean-field approximation (MF2), similar to the method used by Stre\v{c}ka\cite{strecka07} for the quantum Heisenberg TKL model, in which we sum over all eight states of the $a$-trimers with appropriate Boltzmann weights instead of treating each $a$-spin independently.  The mean-field equations are then
 \begin{align}
 m_a &= \frac{  
                         e^{4\beta J_{aa}} \sinh 6\beta J_{ab}m_b
                +        \sinh 2\beta J_{ab}m_b
        }{
                                e^{4\beta J_{aa}} \cosh 6\beta J_{ab}m_b
                +       3 \cosh 2\beta J_{ab}m_b
        },
\nonumber\\
 m_b &= \tanh \beta(4J_{ab}m_a),
 \label{e:mf2}
 \end{align}
and the critical temperature is given by
 \begin{align}
 \frac{T_c^\text{MF2} }{ \left|J_{ab} \right| }
 = \sqrt{\frac{8( 1+3e^{4J_{aa}/T_c^\text{MF2}}  )}{  (3+e^{4J_{aa}/T_c^\text{MF2}}) }}.
 \end{align}

$T_c^\text{MF1}$ and $T_c^\text{MF2}$ are the dotted and dashed curves, respectively, in Fig.~\ref{zerofieldphasediagram}.  The mean-field approximations overestimate $T_c$ by a factor of two or more.  In the regime $J_{aa}/|J_{ab}| \gg 1$, $T_c$ is dominated by the weak links ($J_{ab}$), so $T_c/|J_{ab}|$ tends to a constant, a fact which is captured by MF2.  However, in the frustrated regime $J_{aa}/|J_{ab}| < -1$, both MF1 and MF2 predict the wrong behavior of $T_c$ (see Fig.~\ref{zerofieldphasediagram}).  Also, if an external field is included in the analysis, these mean-field approximations predict an induced ferromagnetic or ferrimagnetic moment, but by construction, they are unable to capture phases III, IV, and the zero-field ``$\ln 72$'' phase in the rich phase diagram presented in Sec.~\ref{zerotemperature}.


\begin{thebibliography}{10}
\newcommand{\enquote}[1]{``#1''}


\bibitem{zhitomirsky-2003}
M.~E. Zhitomirsky, Phys. Rev. B {\bf 67}, 104421 (2003).

\bibitem{zhitomirsky-2004}
M.~E. Zhitomirsky and H.~Tsunetsugu, Phys. Rev. B {\bf 70}, 100403(R) (2004).

\bibitem{derzhko-2004}
O.~Derzhko and J.~Richter, Phys. Rev. B {\bf 70}, 104415 (2004).

\bibitem{isakov-2004}
S.~V. Isakov, K.~S. Raman, R.~Moessner, and S.~L. Sondhi, Phys. Rev. B {\bf
  70}, 104418 (2004).

\bibitem{aoki-2004}
H.~Aoki, T.~Sakakibara, K.~Matsuhira, and Z.~Hiroi, J. Phys. Soc. Japn. {\bf
  73}, 2851 (2004).

\bibitem{gonzalez93}
M.~Gonzalez, F.~Cervantes-Lee, and L.~W. ter Haar, Mol. Cryst. Liq. Cryst. {\bf
  233}, 317 (1993).

\bibitem{maruti94}
S.~Maruti and L.~W. ter Haar, J. Appl. Phys {\bf 75}, 5949 (1993).

\bibitem{mekata98}
M.~Mekata, M.~Abdulla, T.~Asano, H.~Kikuchi, T.~Goto, T.~Morishita, and
  H.~Hori, J. Magn. Magn. Matt. {\bf 177}, 731 (1998).

\bibitem{strecka07}
J.~Stre\v{c}ka, J. Magn. Magn. Matt. {\bf 316}, e346 (2007).

\bibitem{daoxinMSthesis}
D.~X. Yao, {\em Invariant Theory of Time-Dependent Quantum Systems and Ising
  Model on Triangular Kagome Lattice\/}, Master's thesis, Zhejiang University,
  Hangzhou, China (1998).

\bibitem{zheng05}
J.~Zheng and G.~Sun, Phys. Rev. B {\bf 71}, 052408 (2005).

\bibitem{loh2006}
Y.~L. Loh and E.~W. Carlson, Phys. Rev. Lett. {\bf 97}, 227205 (2006).

\bibitem{loh-jeremy2007}
Y.~L. Loh, E.~W. Carlson, and M.~Y.~J. Tan, Phys. Rev. B {\bf 76}, 014404
  (2007).

\bibitem{chineseremainder}
A.~Galluccio, M.~Loebl, and J.~Vondrak, Phys. Rev. Lett. {\bf 84}, 5924 (2000).

\bibitem{wannier50}
G.~H. Wannier, Phys. Rev {\bf 79}, 357 (1950).

\bibitem{kano53}
K.~Kano and S.~Naya, Prog. Theor. Phys. {\bf 10}, 158 (1953).

\bibitem{kasteleyn1963}
P.~W. Kasteleyn, J. Math. Phys. {\bf 4}, 287 (1963).

\bibitem{fisher1966}
M.~E. Fisher, J. Mod. Phys. {\bf 7}, 10 (1966).

\bibitem{horiguchi1992}
T.~Horiguchi, K.~Tanaka, and T.~Morita, J. Phys. Soc. Jpn. {\bf 61}, 64 (1992).

\bibitem{moessner2000}
R.~Moessner, S.~L. Sondhi, and P.~Chandra, Phys. Rev. Lett. {\bf 84}, 4457
  (2000).

\bibitem{moessner2001}
R.~Moessner and S.~L. Sondhi, Phys. Rev. B {\bf 63}, 224401 (2001).

\bibitem{udagawa2002}
M.~Udagawa, M.~Ogata, and Z.~Hiroi, J. Phys. Soc. Jpn. {\bf 71}, 2365 (2002).

\bibitem{moessner2003}
R.~Moessner and S.~L. Sondhi, Phys. Rev. B {\bf 68}, 064411 (2003).

\bibitem{heilman1972}
O.~J. Heilman and E.~H. Lieb, Commun. Math. Phys. {\bf 25}, 190 (1972).

\bibitem{heringa1998}
J.~R. Heringa and H.~W.~J. Bl\"ote, Phys. Rev. E {\bf 57}, 4976 (1998).

\bibitem{krauth-2003}
W.~Krauth and R.~Moessner, Phys. Rev. B {\bf 67}, 064503 (2003).

\bibitem{sandvik-2006}
A.~W. Sandvik and R.~Moessner, Phys. Rev. B {\bf 73}, 144504 (2006).

\bibitem{takagi93}
T.~Takagi and M.~Mekata, J. Phys. Soc. Jpn. {\bf 62}, 3943 (1993).

\bibitem{stephenson64}
J.~Stephenson, J. Math. Phys. {\bf 5}, 1009 (1964).

\bibitem{stephenson70}
J.~Stephenson, J. Math. Phys. {\bf 11}, 413 (1970).

\bibitem{syozi50a}
K.~Husimi and I.~Syozi, Prog. Theor. Phys. {\bf 5}, 177 (1950).

\bibitem{syozi50b}
K.~Husimi and I.~Syozi, Prog. Theor. Phys. {\bf 5}, 341 (1950).

\bibitem{foot1}
The erroneous numerical result in Wannier's original paper ($0.3383 k_B$) is
  often quoted, although it was corrected in an erratum 23 years
  later.~\cite{wanniererratum}.

\bibitem{baxter-1980}
R.~J.~Baxter, J. Phys. A {\bf 13}, L61 (1980).

\bibitem{syozi}
I.~Syozi, \enquote{in {\em Phase Transitions and Critical Phenomena}, Domb and
  Green (eds), vol. 1,}  (1972).

\bibitem{suto1981}
A.~S\"ut\"o, Z. Phys. B {\bf 44}, 121 (1981).

\bibitem{haar95}
S.~M. S.~Ateca and W.~ter Haar, J. Magn. Magn. Matt. {\bf 147}, 398 (1995).

\bibitem{wanniererratum}
G.~H. Wannier, Phys. Rev. B {\bf 7}, 5017 (1973).

\bibitem{fnote}
Note that the enclosed $a$-trimer is {\em not} restricted to a particular $S_z$ subspace,  as shown in Fig.~\ref{zeroTzeroh}(b).

\end{thebibliography}
\bibliographystyle{forprb}

\end{document}